\documentclass[11pt]{article}
\usepackage{graphicx} 
\usepackage{pdflscape} 

\usepackage{geometry}
\usepackage{color}
\usepackage{amsmath,amssymb,amsfonts,}
\usepackage{bm}
\usepackage{booktabs}
\usepackage{enumitem}
\usepackage{hyperref}
\hypersetup{colorlinks=true,linkcolor=black,citecolor=black,urlcolor=blue}
\usepackage{adjustbox}     

\usepackage[x11names, svgnames, rgb]{xcolor}
\usepackage[utf8]{inputenc}
\usepackage{tikz}
\usetikzlibrary{snakes,arrows,shapes}
\usepackage{amsmath}
\usepackage{authblk} 
\usepackage{float}

\usepackage{caption}
\usepackage{fancyvrb}

\usepackage{amsmath,amssymb,amsfonts,mathtools,bm}
\usepackage{microtype}
\usepackage{booktabs}
\usepackage{siunitx}
\usepackage{graphicx}
\usepackage{caption}
\usepackage{hyperref}
\usepackage{natbib}  %
\setcitestyle{authoryear,open={(},close={)}}

\title{Dynamic Attention (DynAttn):  Interpretable High-Dimensional Spatio-Temporal Forecasting\\ (with Application to Conflict Fatalities)}

\author[1]{Stefano M. Iacus}
\author[2,3]{Haodong Qi}
\author[4]{Marcello Carammia}
\author[5]{Thomas Juneau} 

\affil[1]{Harvard University, Institute for Quantitative Social Science, Cambridge (MA), USA.}
\affil[2]{Stockholm University Demography Unit, Stockholm, Sweden}
\affil[3]{Malm\"o University, Department of Global Political Studies, Malm\"o, Sweden}
\affil[4]{University of Catania, Department of Political and Social Sciences, Catania, Italy.}
\affil[5]{University of Toronto, Graduate School of Public and International Affairs, Ottawa, Canada}

\date{}

\begin{document}

\maketitle

\begin{abstract}
Forecasting conflict-related fatalities remains a central challenge in political
science and policy analysis due to the sparse, bursty, and highly non-stationary
nature of violence data. We introduce \textit{DynAttn}, an interpretable
dynamic-attention forecasting framework designed for high-dimensional
spatio-temporal count processes. DynAttn combines rolling-window estimation,
shared elastic-net feature gating, a compact weight-tied self-attention encoder,
and a zero-inflated negative binomial (ZINB) likelihood. This architecture
simultaneously delivers calibrated multi-horizon forecasts of expected casualties
and exceedance probabilities, while retaining transparent diagnostics through
feature gates, ablations, and elasticities.

We evaluate DynAttn using global country-level and high-resolution PRIO-grid-level
conflict data from the VIEWS forecasting system, benchmarking its performance
against established statistical and machine-learning approaches, including
DynENet, LSTM, Prophet, PatchTST, and the official VIEWS baseline. Across horizons
from one to twelve months, DynAttn consistently achieves substantially higher
predictive accuracy, particularly in sparse grid-level settings where alternative
models exhibit severe degradation.

Beyond predictive gains, DynAttn enables structured interpretation of regional
conflict dynamics. Cross-regional analyses show that short-run conflict persistence
and spatial diffusion dominate forecasting performance, while climate stress acts
either as a conditional amplifier or a primary driver depending on the conflict
theater. Slow-moving structural covariates contribute little to short-horizon
forecasting once high-frequency dynamics are accounted for. Taken together, the
results demonstrate that accurate conflict forecasting requires models that are
both probabilistically well-specified and adaptable to heterogeneous regional
regimes. DynAttn provides a flexible and interpretable framework for this task.
\end{abstract}

\clearpage
\tableofcontents

\clearpage

\section{Introduction}
Anticipating armed conflicts and assessing their potential human costs has become increasingly critical in today's global landscape. Accurate forecasts of conflict-related fatalities, along with insights into their underlying drivers, are vital for informing timely interventions, optimizing resource allocation, and mitigating humanitarian crises. Yet, most empirical approaches in social sciences are inadequate in providing reliable predictions and actionable insights \cite{CranmerDesmarais17}. To address this gap, we introduce a novel machine learning framework designed for modeling sparse and noisy count data -- an approach that is particularly well-suited for forecasting conflict-induced casualties and gaining insights of their underlying drivers.

Statistical models in peace and conflict research are generally theoretically derived. The primary objective of these models is to infer how adverse outcomes are associated with different micro and macro factors. The strengths of these associations are quantified by model parameters. To ensure their interpretability, these parameters are typically estimated using (generalized) linear models, e.g., ordinary least square, logistic, probit, or poisson regressions. However, such models might struggle if the dynamics underlying conflict process are nonlinear and shaped by complex interplay of socio-political, economic, and environmental factors. 

Recent advances in computing technology have made it possible to train large and highly complex models, most notably the large language models. While these models can effectively capture nonlinear dynamics, they often operate as a black-box, limiting their utility for explanatory analysis \cite{Haffner_Hofer_Nagl_Walterskirchen_2023}.

To bridge these aforementioned gaps, we introduce DynAttn -- a hybrid modeling framework that integrates basic ideas of  machine learning and modern generative AI architectures with generalized linear models. While this hybrid approach aims to enhance predictive accuracy of conflict deaths, it also provides mechanisms to gain explanatory insights and hence ensuring interpretability. 

We evaluate DynAttn against established forecasting methods, including Prophet, Long Short-Term Memory (LSTM) networks, dynamic elastic-net, PatchTST, using the comprehensive dataset provided by the Violence Early-Warning System (VIEWS) \citep{Hegre2019VIEWS, Hegre2021VIEWS2020}. Our results demonstrate that DynAttn improves predictive accuracy while maintaining interpretability and offering insights into how conflict-induced mortality risks may evolve over time and space. 

By situating complex neural networks within tractable and rigorous statistical models, this paper contributes to the methodological innovation in political science and beyond. We demonstrate how hybrid models can reconcile the trade-off between complexity and interpretability in forecasting tasks, advancing both the empirical study of conflict dynamics and the broader integration of computational methods in political analysis.

The remainder of the paper is organized as follows. Section~\ref{sec:method}
introduces the DynAttn framework for nowcasting conflict intensity. In addition to
presenting the model architecture in detail, this section situates DynAttn within
the existing conflict forecasting literature by briefly reviewing pre-transformer
and machine-learning approaches, thereby clarifying the methodological
contributions of the proposed framework. It then discusses key modeling choices
and implementation decisions, reports the resulting parameter count, and outlines
possible extensions. Section~\ref{sec:data} describes the data sources, outcome
variables, and covariates used in the analysis.
Section~\ref{sec:perf} evaluates the predictive performance of DynAttn. After describing the
validation setup, we assess forecasting accuracy for casualties using
correlation and prediction error metrics, examine the model’s ability to predict
conflict risk, and summarize overall performance relative to alternative
approaches. The section concludes with an analysis of forecast explainability,
using ablation importance and elasticity measures.

Building on these diagnostics, Section~\ref{sec:understand} uses the model’s gating and
spatio-temporal structure to examine conflict dynamics at high spatial
resolution. We analyze four regional conflict systems: Central/East Africa,
the Middle East, the Sahel, and the Horn of Africa, and highlight how different
theaters exhibit distinct forecasting regimes characterized by varying
combinations of temporal persistence, spatial diffusion, and environmental
stress. This section ends with a cross-regional synthesis of common structures
and regional specificities. Section~\ref{sec:conclusions} concludes.

\section{Nowcasting Conflict Intensity via DynAttnn}
\label{sec:method}

A natural starting point for our work is the \textbf{Dynamic Elastic Net (DynENet)} 
framework, originally introduced to forecast migration flows using large-scale, 
heterogeneous data sources—including official statistics, operational data, and 
social media \citep{carammia22}. DynENet extends the classical elastic net algorithm 
by training over a \emph{rolling sequence} of windows, which makes the linear model 
adaptive to evolving data regimes and mitigates non-stationarity. This rolling 
training phase selects the best specification for subsequent forecasting, providing 
a parsimonious and interpretable baseline through sparse linear relationships.

Despite its recent introduction, DynENet has already been applied in multiple domains.
In political science, it has been used to predict conflict-related fatalities—an 
outcome characterized by burstiness and irregular temporal patterns \citep{iacus2024conflict}. 
It has also been used in forecasting asylum-related migration flows \citep{carammia22} 
and in climate-induced migration, where covariates are high-dimensional and dynamic 
\citep{Haodong25}. These applications highlight the value of dynamic feature selection 
when dealing with complex and evolving signal environments.

However, DynENet inherits key limitations from linear models.  
First, it assumes Gaussian residuals\footnote{
Extensions to non-Gaussian likelihoods exist, but rely on the same optimization 
structure, which becomes difficult to fit for zero-inflated and over-dispersed 
counts such as those considered here.}  
and minimizes a squared-error objective. These assumptions may be appropriate 
for continuous or dense outcomes, but they are inadequate for discrete, sparse, 
and heavy-tailed conflict fatalities. Gaussian-based formulations underestimate 
tail risks and cannot coherently model exceedance probabilities 
(e.g., $\Pr(Y\!\ge\!\tau)$).

To address these limitations, we introduce \textbf{DynAttnn}, a modeling framework 
designed for sparse, heavy-tailed, and risk-focused forecasting problems. DynAttnn 
replaces the Gaussian residual model with a \textbf{Zero-Inflated Negative Binomial 
(ZINB)} likelihood, enabling principled modeling of over-dispersion and structural 
zeros while providing calibrated exceedance probabilities from a \emph{single 
probabilistic head}. DynAttnn also incorporates a compact self-attention encoder 
and a multi-horizon decoding head within a deep neural architecture, providing 
nonlinear interactions while retaining transparent feature selection through 
shared elastic-net gates.

\subsection{Transformers for Time Series Analysis}

Recent advances in deep learning—driven largely by natural language processing—
have been shaped by the emergence of the \textbf{Transformer} architecture 
\citep{vaswani2017attention}. Transformers replaced recurrent units with 
\emph{self-attention}, enabling parallel training, flexible receptive fields, 
and efficient modeling of long-range dependencies. Their success led to the 
development of large pretrained language models (LLMs) such as BERT 
\citep{devlin2019bert} and GPT-family models \citep{radford2019language,brown2020language}, 
which now form the foundation for a wide range of tasks across domains.

For time-series forecasting, the Transformer family offers two opportunities 
especially relevant to conflict analysis:
(i) a compact mechanism to encode nonlinear interactions across recent time 
windows of high-dimensional covariates; and 
(ii) the ability to combine \emph{probabilistic fidelity} (via appropriate 
likelihoods) with \emph{interpretability} (via feature selection and attribution).

A variety of Transformer-based models have been proposed for time series. 
\textbf{Informer} \citep{zhou2021informer} introduces probabilistic sparse 
attention for handling long sequences efficiently. \textbf{Autoformer} 
\citep{wu2021autoformer} decomposes inputs into trend and seasonal components 
coupled with autocorrelation attention. \textbf{FEDformer} \citep{zhou2022fedformer} 
moves attention into the frequency domain. \textbf{PatchTST} \citep{nie2023patchtst} 
tokenizes subseries into patches and uses channel-independent encoders, achieving 
strong accuracy on standard benchmarks. \textbf{TimesNet} \citep{wu2023timesnet} 
maps 1-D sequences into learned 2-D temporal structures. Finally, the 
\textbf{Temporal Fusion Transformer (TFT)} \citep{lim2021temporal} combines 
recurrent components with attention and includes variable-selection layers for 
interpretability.

Despite these advances, most Transformer-based time-series models optimize MSE 
or quantile losses under Gaussian-like assumptions. Explicit count distributions—
especially those with zero inflation—remain rare. Probabilistic deep-learning 
baselines such as \textbf{DeepAR} \citep{salinas2020deepar} incorporate Poisson 
or NB likelihoods, but do not support zero inflation and provide limited 
interpretability in high-dimensional settings.

\bigskip

\noindent\textbf{DynAttnn} bridges these literatures by integrating:
\begin{itemize}[leftmargin=2em]
\item[(i)] \textbf{rolling windows} and \textbf{shared elastic-net feature gates} 
for parsimony, stability, and diagnostic interpretability (as in DynENet),
\item[(ii)] a compact, weight-tied \textbf{self-attention encoder} summarizing 
nonlinear interactions across the previous $S$ months,
\item[(iii)] a \textbf{multi-horizon decoder} that produces $H$ horizon-specific 
embeddings from a single pooled representation, and
\item[(iv)] a \textbf{ZINB probabilistic head} that outputs coherent expected 
counts and exceedance probabilities for all horizons.
\end{itemize}

To our knowledge, explicit ZINB likelihoods combined with feature-gated attention 
inside a rolling multi-horizon nowcasting pipeline are uncommon in the forecasting 
literature. Table~\ref{tab:DynAttnn_compare} positions DynAttnn relative to the 
most relevant alternatives.

\begin{table}[t]
\centering
\caption{Positioning DynAttnn among econometric and Transformer-based forecasters.}
\label{tab:DynAttnn_compare}
{\tiny
\begin{tabular}{p{2.9cm} p{3.7cm} p{3.7cm} p{3.7cm}}
\toprule
\textbf{Aspect} &
\textbf{DynENet} &
\textbf{Transformer models} \\
& (rolling ENet) &
(Informer, Autoformer, FEDformer, PatchTST, TFT, TimesNet) &
\textbf{DynAttnn} \\
\midrule
\textbf{Temporal handling} &
Rolling windows; linear dynamics &
Deep encoder stacks or patch tokenization; long-horizon focus &
Rolling windows with compact, weight-tied attention over last $S$ months \\

\addlinespace[2pt]
\textbf{Exogenous covariates} &
High-dimensional linear EN &
High-dimensional nonlinear embeddings; limited built-in sparsity &
High-dimensional, \emph{shared feature gates} for sparsity + stability \\

\addlinespace[2pt]
\textbf{Likelihood / loss} &
Gaussian / squared error &
Mostly MSE or quantile losses; count heads uncommon &
\textbf{ZINB} (over-dispersion + structural zeros); calibrated $\Pr(Y\!\ge\!\tau)$ \\

\addlinespace[2pt]
\textbf{Interpretability} &
Sparse coefficients &
Attention maps, occasional variable selection (e.g., TFT) &
Gates + ablation + elasticity \\

\addlinespace[2pt]
\textbf{Sample regime} &
Strong with short histories and many regressors &
Often data-hungry; best on long benchmarks &
Optimized for modest-length, region-specific, sparse conflict series \\

\addlinespace[2pt]
\textbf{Typical use cases} &
Multi-regressor linear nowcasting &
Benchmark forecasting (electricity, traffic, ETTh/ETTm datasets) &
Risk-focused nowcasting with exceedance probabilities \\
\bottomrule
\end{tabular}
}
\end{table}

\subsection{Comparison with Pre-Transformer Models}

Traditional econometric approaches rely heavily on assumptions of stationarity 
and Gaussian residuals. We briefly compare DynAttnn with representative methods.

\begin{enumerate}[leftmargin=1.5em]

\item \textbf{DynAttnn vs DynENet.}
\begin{itemize}[leftmargin=0.6em]
\item[] \emph{Distributional fidelity.}  
DynENet targets conditional means; DynAttnn models over-dispersion and zero inflation 
and yields coherent exceedance probabilities.
\item[] \emph{Nonlinear interactions.}  
DynAttnn uses weight-tied attention to model nonlinear structure at low variance.
\item[] \emph{Feature selection.}  
DynAttnn uses a \emph{shared} gate vector across horizons, improving stability.
\item[] \emph{When DynENet may win.}  
In very short histories or purely linear settings.
\end{itemize}

\item \textbf{DynAttnn vs LSTM.}
\begin{itemize}[leftmargin=0.6em]
\item[] \emph{Sample efficiency.}  
LSTMs excel with long sequences; DynAttnn is optimized for short windows with rich covariates.
\item[] \emph{Interpretability.}  
DynAttnn offers gates, ablations, and elasticities; LSTMs are opaque.
\item[] \emph{Likelihood.}  
DynAttnn natively models zero-inflated counts and exceedance probabilities.
\end{itemize}

\item \textbf{DynAttnn vs Prophet.}
\begin{itemize}[leftmargin=0.6em]
\item[] \emph{Non-seasonal, intermittent counts.}  
Prophet assumes seasonal structure; conflict counts are bursty and non-seasonal.
\item[] \emph{Exogenous signals.}  
Prophet handles few regressors; DynAttnn handles hundreds.
\item[] \emph{Tail risks.}  
Prophet does not estimate $\Pr(Y\!\ge\!\tau)$; DynAttnn does.
\end{itemize}

\item \textbf{DynAttnn vs ARIMA/ARIMAX/ETS.}
\begin{itemize}[leftmargin=0.6em]
\item[] \emph{Nonlinear multivariate structure.}  
ARIMA cannot leverage high-dimensional nonlinear covariates.
\item[] \emph{Distribution mismatch.}  
Gaussian residuals miss zero inflation and heavy tails.
\item[] \emph{When classical models may win.}  
Smooth univariate series with few covariates.
\end{itemize}

\end{enumerate}

\begin{table}[t]
\centering
\caption{Conceptual comparison in the conflict nowcasting setting.}
\label{tab:compare}
{\tiny
\begin{tabular}{p{3cm}p{2.6cm}p{2.6cm}p{2.6cm}p{2.6cm}p{2.6cm}}
\toprule
\textbf{Aspect} & 
\textbf{DynENet} & 
\textbf{LSTM} & 
\textbf{Prophet} & 
\textbf{PatchTST} & 
\textbf{DynAttnn} \\
\midrule

\textbf{Observational model} &
Linear, SE loss &
Usually Gaussian head &
Gaussian residuals &
MSE/quantile; no count likelihood &
ZINB (over-dispersion + zeros) \\

\textbf{Exceedance probability} &
Not native &
Not native &
Not native &
Not native &
Directly from ZINB survival \\

\textbf{Exogenous covariates} &
Many, linear EN &
Many, nonlinear &
Few, linear &
Multivariate, channel-independent embeddings; no sparsity &
Many, gated + nonlinear \\

\textbf{Temporal modeling} &
Rolling linear windows &
Recurrent sequence model &
Trend/seasonal decomposition &
Patch tokenization + self-attention encoder &
Windowed, weight-tied self-attention + multi-horizon decoder \\

\textbf{Interpretability} &
Sparse coefficients &
Opaque &
Additive components &
Limited (attention maps only) &
Gates + ablation + elasticity \\

\textbf{Best regime} &
Short, simple links &
Long sequences, complex motifs &
Seasonal business series &
Long-history multivariate datasets; strong on benchmarks &
Zero-inflated, bursty counts with rich covariates \\
\bottomrule
\end{tabular}
}
\end{table}

\subsection{Key Advantages of DynAttnn in Forecasting Conflict Intensity}

\begin{enumerate}[leftmargin=1.4em]
\item \textbf{Calibrated multi-horizon probabilities from a single model.}  
The ZINB head produces both $\widehat{y}_{t,h}$ and $\Pr(Y_{t+h}\!\ge\!\tau)$ coherently for all horizons.

\item \textbf{Correct likelihood for sparse, bursty counts.}  
Zero inflation and over-dispersion are modeled explicitly.

\item \textbf{Scalable feature selection with diagnostics.}  
A single shared gate vector stabilizes selection across horizons; ablation and elasticity 
quantify predictor salience.

\item \textbf{Adaptive forecasting without recursion.}  
Rolling training captures regime changes; forecasts are \emph{direct} rather than recursively unrolled.

\item \textbf{Sample-efficient attention.}  
A shallow, windowed, weight-tied encoder captures nonlinear signals with fewer parameters 
than recurrent stacks.
\end{enumerate}

\subsection{DynAttn Model description}

\begin{figure}[t]
  \caption{Architecture of the gated attention ZINB model.}
  \label{fig:flowchart}
  \centering
  \resizebox{0.9\textwidth}{!}{\includegraphics{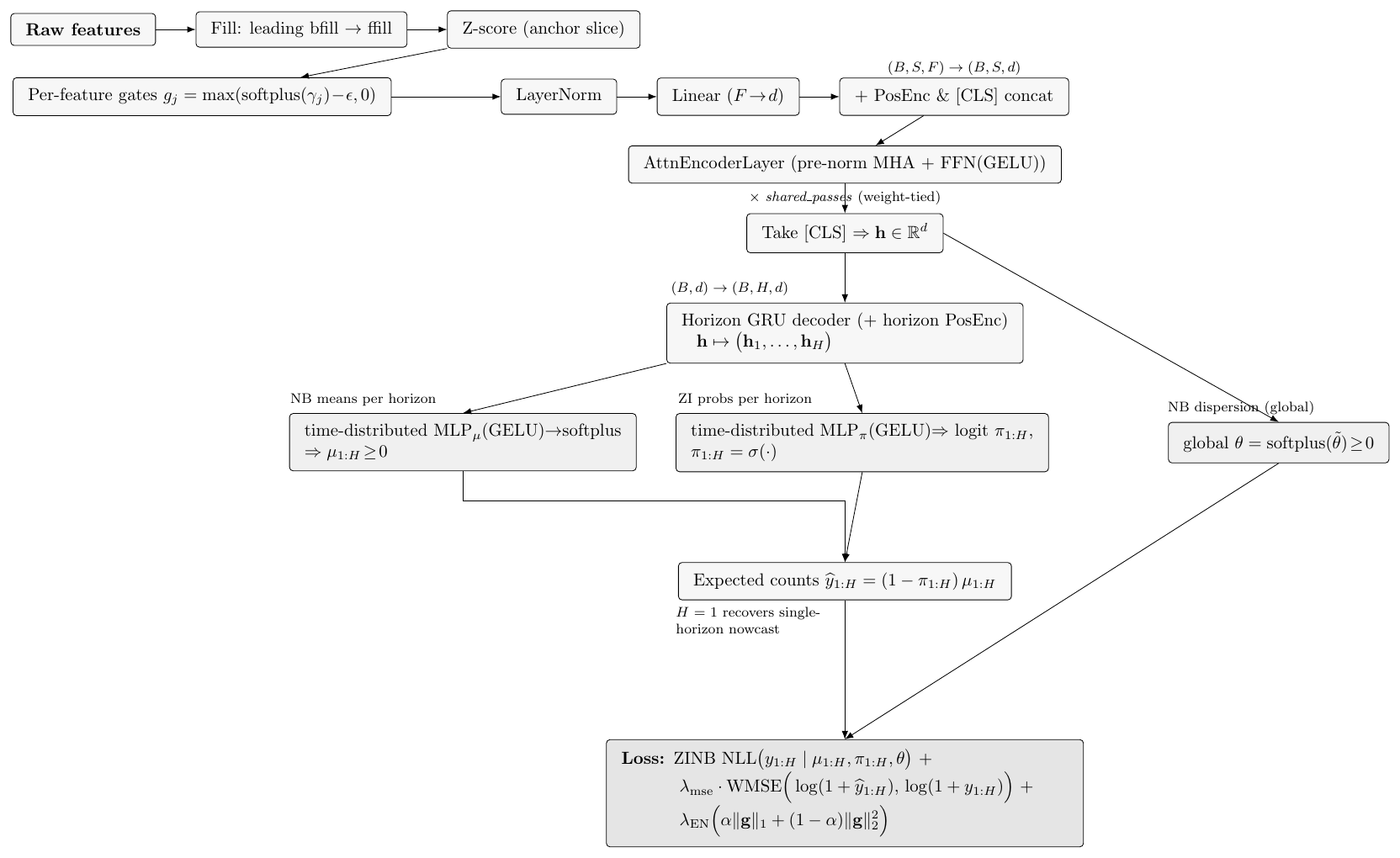}}
\end{figure}

Figure~\ref{fig:flowchart} illustrates the architecture of the DynAttnn model used in this study.  
The objective is to \emph{forecast monthly conflict intensity}, defined as the number of casualties recorded in future months. For each time $t$ and vector of forecast horizons $H=\{1, 2, \ldots, H\}$, the model target is the vector of forecasts
\[
(\text{CAS}_{t+1}, \text{CAS}_{t+2}, \ldots, \text{CAS}_{t+H} )
\]
with each $\text{CAS}_{t+h}\ge 0$, 
 representing the casualty count observed in month $t+h$, $h\in H$.  
From the same probabilistic output layer the model also returns a horizon specific exceedance probability,
\[
\text{PR}_{t,h} = \Pr\!\big( \text{CAS}_{t+h} \ge \tau \big),
\]
with thresholds $\tau$. In the application below, $\tau  = 25$ for the country level forecasts and $\tau = 1$ for the PRIO grid level forecasts.

The model combines three components:  
(i) non-negative \textit{feature gates} with elastic net regularization, which identify the most relevant predictors while improving stability in this high dimensional setting;  
(ii) a compact \textit{self attention encoder} that summarizes the last $S$ months of standardized inputs; and  
(iii) a \textit{zero inflated negative binomial} (ZINB) output layer that captures both over dispersion and the large fraction of zero casualty months.

\paragraph{Windows, standardization, and gating.}

Let $\mathbf{X}\in\mathbb{R}^{T\times p}$ denote the matrix of $p$ raw predictors used to forecast the monthly outcome $\text{CAS}_{t+H}$.  
For each anchor month $t$ the model constructs a fixed length window
\[
\mathbf{X}_t = [\,\tilde{\bm x}_{t-S+1},\ldots,\tilde{\bm x}_{t}\,] \in \mathbb{R}^{S\times p},
\]
where $\tilde{\bm x}_s$ denotes the vector of predictors at time $s$ after pre-processing.  
Missing values, including those arising from lagged features, are imputed by back filling from the nearest previous non missing value and then forward filling; any remaining non finite entries are set to zero.

To prevent target leakage, standardization is performed conditionally on the information available up to the anchor.  
For each predictor $j$, the mean $\mu_{t,j}$ and standard deviation $\sigma_{t,j}$ are computed using the history up to $t$, and all entries of the window are standardized using these parameters.  
This ensures that both the model and the standardization logic use only past information at every prediction date.

Feature selection is implemented through a nonnegative gate vector $\bm g \in \mathbb{R}^p_{\ge 0}$ that rescales predictors featurewise,
\[
\mathbf{X}_t^{(g)} = \mathbf{X}_t \operatorname{diag}(\bm g),
\qquad
g_j = \max\!\big(\mathrm{softplus}(\gamma_j) - \epsilon,\; 0\big),
\]
where $\gamma_j\in\mathbb{R}$ are unconstrained parameters and $\epsilon = 10^{-4}$.  
The gate vector is penalized through an elastic net term
\[
\lambda_{\mathrm{EN}}\big( \alpha \|\bm g\|_1 + (1-\alpha)\|\bm g\|_2^2 \big),
\]
which promotes sparse and stable use of predictors.  
Variables with the largest estimated gates $g_j$ are interpreted as selected features, for example using a predefined magnitude threshold or by retaining the top fraction of nonzero gates.

\paragraph{Transformer encoder and pooled representation.}

After gating and standardization, each window $\mathbf{X}_t^{(g)} \in \mathbb{R}^{S\times p}$ is mapped to a sequence of
$S$ vectors through a linear projection into a $d$-dimensional hidden space.  
A special learnable vector, denoted \texttt{[CLS]}, is prepended to the sequence.  
In the language of natural language processing this behaves like a token, although it does not correspond to any real input.  
The purpose of \texttt{[CLS]} is to serve as a summary slot.  
During training it learns to attend to the full sequence, and the sequence learns to attend back to it.  
In practice the \texttt{[CLS]} state acts as a pooled representation of the entire window.\footnote{
In NLP tasks sentences are decomposed into tokens and Transformers operate on fixed-length sequences of embeddings.  
The \texttt{[CLS]} token is a learnable vector added to the sequence so that the attention mechanism has a designated location that integrates information over all positions.  
Its final hidden state is used as a summary representation of the whole input.
}

Position-specific information is added through sinusoidal positional encodings
applied to the projected feature vectors.  
The augmented sequence
\[
\big[\,\texttt{[CLS]},\;\mathbf{X}_t^{(g)} W + \mathrm{PosEnc}\,\big]
\]
is then passed through $L$ weight-tied self-attention blocks.  
Weight tying means that a single encoder layer is applied repeatedly $L$ times, which yields a deep implicit receptive field while keeping the number of parameters fixed.

Let $\mathrm{AttnEnc}(\cdot)$ denote these $L$ applications of the pre-norm multihead self-attention block.  
The pooled representation used by the likelihood is the terminal \texttt{[CLS]} state,
\[
\mathbf{h}_t
\;=\;
\mathrm{AttnEnc}\!\left(
 \big[\texttt{[CLS]},\,\mathbf{X}_t^{(g)}W + \mathrm{PosEnc}\big]
\right)_{\texttt{[CLS]}}
\;\in\;\mathbb{R}^d.
\]

\paragraph{MLP heads and GELU activations.}

The pooled vector $\mathbf{h}_t$ is passed to two shallow feedforward heads, both using Gaussian Error Linear Units (GELU) \citep{hendrycks2016gelu} as the nonlinear activation.  
GELU improves numerical stability when many inputs are zero, a frequent situation in violence count data:
\[
\mathrm{GELU}(x)=x\cdot\Phi(x),
\qquad
\Phi(x)\text{ is the standard normal cdf}.
\]
Compared to $\mathrm{ReLU}(x)=\max\{0,x\}$, GELU provides smooth gating and avoids hard cutoffs.
Each head is a two-layer MLP  (Multi-Layer Perceptron):
\[
\mathrm{MLP}(x)=W_2\,\mathrm{GELU}(W_1 x + b_1)+b_2.
\]
From $\mathbf{h}_t$ we compute:
\[
\begin{aligned}
\tilde{\mu}_t &= \mathrm{MLP}_{\mu}(\mathbf{h}_t),\\
\tilde{\pi}_t &= \mathrm{MLP}_{\pi}(\mathbf{h}_t).
\end{aligned}
\]
These are mapped to valid Zero-Inflated Negative Binomial (ZINB) parameters:
\[
\mu_t=\mathrm{softplus}(\tilde{\mu}_t)\ge 0,\qquad
\pi_t=\sigma(\tilde{\pi}_t)\in(0,1),\qquad
\theta=\mathrm{softplus}(\tilde{\theta})\ge 0.
\]
The dispersion parameter $\theta$ is global: it is estimated once and shared across all anchors and windows.  
This allows the model to accumulate information about overdispersion across time.

\paragraph{ZINB likelihood, expected counts, and exceedance probability.}

Conditional on $(\mu_t,\pi_t,\theta)$, the ZINB model specifies
\[
\Pr(Y_t=0)=\pi_t+(1-\pi_t)\Pr_{\mathrm{NB}}(0;\mu_t,\theta),\qquad
\Pr(Y_t=y>0)=(1-\pi_t)\Pr_{\mathrm{NB}}(y;\mu_t,\theta).
\]
The expected count used for training, evaluation, and forecasting is
\[
\widehat{y}_t=\mathbb{E}[Y_t]=(1-\pi_t)\mu_t.
\]
For exceedance prediction, the model outputs a calibrated probability that monthly violence exceeds a threshold $\tau$,
\[
\widehat{\mathrm{pr}H_t}
=
\Pr(Y_t\ge \tau)
=
(1-\pi_t)\,\mathrm{NB}\text{-SF}(\tau;\mu_t,\theta),
\]
where the survival function is computed either through the regularized incomplete beta function or through a stable summation procedure when required.

\paragraph{Loss function.}
The overall loss combines the ZINB negative log-likelihood, an auxiliary weighted mean squared error in log space, and the elastic-net penalty on the gates:
\[
\begin{aligned}
\mathcal{L} \;=\;&
\frac{1}{|\mathcal{B}|}\sum_{t\in\mathcal{B}}
\ell_{\mathrm{ZINB}}(y_t;\mu_t,\theta,\pi_t)
\;+\;
\lambda_{\mathrm{mse}}
\frac{1}{|\mathcal{B}|}\sum_{t\in\mathcal{B}} w_t
\bigl(\log(1+\widehat{y}_t)-\log(1+y_t)\bigr)^2
\\[4pt]
&+
\lambda_{\mathrm{EN}}
\bigl(\alpha\|\bm g\|_1+(1-\alpha)\|\bm g\|_2^2\bigr),
\end{aligned}
\]
with
\[
w_t \;=\; 1 + \alpha_{\mathrm{wmse}}\,\mathbf{1}\{y_t>0\}.
\]
Here $\lambda_{\mathrm{mse}}\ge 0$ controls the strength of the auxiliary weighted MSE term and $\alpha_{\mathrm{wmse}}\ge 0$ controls how much additional weight is given to nonzero casualty observations.  
In our experiments we set $\alpha_{\mathrm{wmse}}=3$, which reduces the dominance of zeros and encourages the model to fit the tail of the distribution.

\paragraph{Rolling optimization and region-specific models.}

The model is trained using rolling windows over monthly anchors.  
At each anchor the optimizer state is carried forward, which produces cumulative adaptation over time.  
For computational control an optional cap on the most recent $K$ windows is available.  
Mixed-precision arithmetic (CUDA bf16 or fp16, and MPS fp16) is used whenever available.  
A deterministic FP32 mode disables AMP and TF32 for exact reproducibility.

The entire procedure is run separately for each spatial unit (country or PRIO grid), exactly as in \citet{iacus2024conflict}.  
This produces models that adapt not only over time but also across space, with each region learning its own feature gates and temporal dynamics.

\subsection{Modeling Choices and Implementation}

The introduction of elastic-net gates $\bm g$ provides a principled mechanism for interpretability.  
Each gate controls the effective contribution of a predictor across all time windows, and the learned values can be inspected directly.  
The model outputs support two forms of explainability:  
(i) \emph{elasticities}, which quantify how the forecasted mean changes when a single predictor is perturbed by a fixed percentage (for example, a 10\% increase); and  
(ii) \emph{ablations}, which measure the absolute and relative change in MSE when a predictor is selectively removed by forcing its gate to zero.  
Ablations use a fixed standardization computed on the entire series for numerical stability, whereas elasticities reuse the per-anchor standardization that is applied during training, ensuring that perturbations reflect the model's operational feature space.

A key design choice is that the gate vector $\bm g$ is shared across all forecast horizons in the multi-horizon decoder.  
This constraint reduces variance, improves stability, and enhances interpretability: the same predictors influence all horizons, while horizon-specific dynamics are handled by the decoder.  
Allowing each horizon to learn a separate gate vector $\bm g_h$ leads to severe overparameterization in sparse, region-specific time series and causes gate drift, which in turn degrades elasticity and ablation diagnostics.  
A single shared $\bm g$ acts as a global relevance filter applied before temporal processing, ensuring that horizon-specific behaviour is captured through the recurrent decoder rather than through redundant feature selection layers.

\paragraph{Softplus gates for stable and interpretable feature selection.}

The gate vector $\bm g$ is intended to provide both sparsity and interpretability, which requires a smooth and stable parameterization in sparse, zero-heavy forecasting regimes.  
To achieve this, we parameterize each gate as
\[
g_j = \max\bigl(\mathrm{softplus}(\gamma_j)-\epsilon,\,0\bigr),
\qquad \epsilon = 10^{-4},
\]
with unconstrained $\gamma_j\in\mathbb{R}$.  
This construction has several advantages.  
First, the softplus transformation maintains strictly positive derivatives even when $\gamma_j$ is near or below zero, which prevents gates from becoming permanently inactive when gradients are small.  
Second, the combination of a small subtractive offset $\epsilon$ and the outer maximum operator permits exact zeros when supported by the data, which is essential for interpretability and for enforcing true sparsity.  
Third, the smooth curvature of softplus avoids the nondifferentiability at zero that can cause optimizer instability under mixed precision.  
In practice, this parameterization produces gates that remain responsive to new evidence while avoiding an accumulation of tiny but noninformative values, leading to more reliable feature selection in regions with limited data and many zero observations.

\paragraph{GELU in the encoder and heads vs.\ ReLU.}

The encoder and all MLP heads use the Gaussian Error Linear Unit,
\[
\mathrm{GELU}(x) = x \cdot \Phi(x),
\]
where $\Phi$ is the standard normal cdf.  
GELU retains small but nonzero responses near zero and has smoother curvature than ReLU.  
This is advantageous because many standardized covariates remain close to zero for extended periods (for example, during quiescent months in conflict series).  
ReLU would clip these weak signals entirely, yielding brittle gradients.  
GELU, in contrast, produces smoother optimization dynamics, improves calibration of the ZINB parameters, and interacts more reliably with automatic mixed precision (AMP).  
These benefits are particularly important when the window length is relatively short ($S \approx 48$), when data are region-specific and limited, and when gates plus elastic-net regularization control variance rather than deep stacks of independent parameters.

\paragraph{Softplus links for $\mu$ and $\theta$.}

The NB mean $\mu$ and the dispersion parameter $\theta$ must be nonnegative.  
Mapping the outputs of the MLP heads through $\mathrm{softplus}(\cdot)$ ensures this constraint while maintaining smooth gradients.  
Softplus stabilizes the term $\log(\theta+\mu)$ in the ZINB likelihood and prevents negative or exactly zero $\theta$, which would cause the log-likelihood to diverge.  
ReLU, by contrast, would produce large zero plateaus and kinks that behave poorly under AMP.  
Softplus provides well-behaved curvature in the region where most of the probability mass lies: small $\mu$ with many zeros.

\paragraph{Sigmoid link for zero inflation $\pi$.}

The zero-inflation probability $\pi$ must lie in $(0,1)$, and we therefore parameterize it through logits using the logistic sigmoid.  
This choice has two advantages:  
(i) logits combine additively with the NB zero mass on the log scale in a numerically stable manner; and  
(ii) the model can increase $\pi$ in response to extended runs of zeros without forcing $\mu\to 0$, which would collapse the NB component and degrade tail calibration.  
Maintaining a separation between the structural-zero channel ($\pi$) and the NB intensity channel ($\mu$) is critical for accurate exceedance probabilities $\Pr(Y\ge \tau)$ in bursty time series.

\paragraph{Why a global dispersion $\theta$.}

The model estimates a single global dispersion parameter $\theta$ shared across all windows for a given region.  
Although a time-varying sequence $\theta_t$ could in principle capture local changes in over-dispersion, it becomes highly unstable when positive observations are rare, causing erratic gradients in the survival function and undermining both $\widehat{y}_t$ and $\widehat{\mathrm{pr}H_t}$.  
A global $\theta$ regularizes the NB shape and improves the stability of both the expected count and the exceedance probability, especially at longer horizons.

\paragraph{Why the auxiliary WMSE in $\log(1+\cdot)$.}

The ZINB log-likelihood models the full count distribution, but we supplement it with a weighted MSE on the expected count in log1p space,
\[
\bigl(\log(1+\widehat{y}_t) - \log(1+y_t)\bigr)^2,
\]
which receives additional weight when $y_t>0$ through
\[
w_t = 1 + \alpha_{\mathrm{wmse}}\, \mathbf{1}\{y_t>0\}.
\]
This auxiliary loss improves calibration of the conditional mean while preserving attention to rare positive outcomes.  
The $\log(1+\cdot)$ transform compresses heavy tails and stabilizes gradients when spikes occur.  
In practice, $\alpha_{\mathrm{wmse}}$ is set to $3$ in the multi-horizon experiments.

\paragraph{Why shallow, weight-tied attention.}

The encoder repeatedly applies a single self-attention layer for $L$ passes with shared weights.  
This captures within-window interactions while constraining the parameter count, which is beneficial in region-specific settings with limited data.  
Weight tying reduces variance and prevents overfitting compared to deeper stacks of independently parameterized layers.  
The learned \texttt{[CLS]} vector acts as a dedicated summary token that collects information across months and provides a compact representation suitable for parameterizing the ZINB output.

\subsection{Parameter Count}

The number of trainable parameters in the \textbf{DynAttnn} model can be computed in closed form.
Let $F$ denote the number of predictors after preprocessing, $d=256$ the hidden size of the encoder, and $h=128$ the hidden width of each MLP head.  
The total parameter count decomposes as follows:
\[
\begin{aligned}
N_{\text{params}}
\;=\;&
\underbrace{F}_{\text{gates}}
\;+\;
\underbrace{2F}_{\text{input norm}}
\;+\;
\underbrace{Fd + d}_{\text{input projection}}
\;+\;
\underbrace{d}_{\texttt{[CLS] token}}
\;+\;
\underbrace{(12d^2 + 13d)}_{\text{weight\text{-}tied attention encoder}}
\\[6pt]
&\;+\;
\underbrace{2(dh + 2h + 1)}_{\text{two MLP heads}}
\;+\;
\underbrace{1}_{\theta\text{ (dispersion)}}.
\end{aligned}
\]
Collecting terms yields the closed-form expression
\[
N_{\text{params}} = 259F + 856{,}323.
\]

Table~\ref{tab:paramcount} reports the resulting parameter counts for representative feature sizes.  
In our empirical application the country-level model uses $F=123$ predictors and the PRIO grid model uses $F=104$.  
The attention encoder accounts for the majority of the parameters, reflecting the fact that the dominant computational footprint is driven by the embedding dimension $d$ rather than by the number of predictors $F$.  
As a consequence, the model scales gracefully as additional exogenous features are incorporated: increasing $F$ primarily affects the small linear projection layer and the gates, while the encoder cost remains fixed.

\begin{table}[h!]
\centering
\caption{Approximate parameter counts for different feature dimensions $F$.}
\label{tab:paramcount}
\begin{tabular}{l r}
\toprule
Feature count $F$ & Total parameters $N_{\text{params}}$ \\
\midrule
104 (PRIO grid level) & $883{,}259$ \\
123 (country level) & $888{,}180$ \\
300 & $933{,}023$ \\
500 & $985{,}823$ \\
\bottomrule
\end{tabular}
\end{table}

Overall, the \textbf{DynAttnn} architecture remains compact, typically below one million parameters even when hundreds of exogenous features are included.  
This compactness is crucial for region-specific forecasting, where the amount of training data per model is limited and efficient learning requires a strong inductive bias together with a constrained parameter footprint.

\subsection{Model Extensions}
\label{sec:extensions}

Although our empirical application focuses on zero-inflated count outcomes via a 
ZINB likelihood, the DynAttn architecture is designed as a modular 
forecasting framework capable of supporting a broader family of observation models. 
This modularity isolates the distribution-specific components from the dynamic 
attention encoder, the shared feature-gating mechanism, and the multi-horizon 
decoder. In this section, we outline the likelihood families currently supported 
and clarify which parts of the architecture are affected by such extensions.

\paragraph{General structure of the likelihood head.}
All likelihoods in DynAttn share the same upstream components:
\begin{enumerate}[label=(\roman*), itemsep=2pt, topsep=2pt]
\item window construction and dynamic standardization,
\item shared softplus feature gates,
\item the weight-tied attention encoder producing a pooled state $\mathbf{h}_t$,
\item the multi-horizon decoder mapping $\mathbf{h}_t \mapsto \{\mathbf{h}_{t,h}\}_{h=1}^H$.
\end{enumerate}
\noindent
\textit{Only} the final distributional layer, the likelihood head, changes when 
switching to a different observation model. This separation makes DynAttn both 
flexible and extensible, while ensuring that meaningful extensions require careful 
statistical and architectural adjustments.

\subsubsection*{Currently supported likelihood families in the software implementation}

DynAttn currently implements several observation models, each appropriate for 
different data-generating processes.

\paragraph{Gaussian likelihood (continuous outcomes).}
For symmetric, continuous targets,
\[
Y_{t+h} \sim \mathcal{N}(\mu_{t,h},\,\sigma^2_{t,h}),
\]
where the MLP head outputs unconstrained parameters
$\tilde{\mu}_{t,h}$ and $\tilde{\sigma}_{t,h}$, followed by
\[
\mu_{t,h} = \tilde{\mu}_{t,h}, 
\qquad
\sigma_{t,h} = \mathrm{softplus}(\tilde{\sigma}_{t,h}).
\]
Only the likelihood head and its NLL change; the encoder and gating structure 
remain unchanged.

\paragraph{Student-$t$ likelihood (robust continuous modeling).}
For heavy-tailed continuous observations,
\[
Y_{t+h} \sim t_{\nu}(\mu_{t,h},\,\sigma_{t,h}),
\]
with $\nu$ fixed or learned. This likelihood provides robustness to extreme 
values at the cost of a more complex negative log-likelihood.

\paragraph{Poisson likelihood (non-overdispersed counts).}
For count data without over-dispersion,
\[
Y_{t+h} \sim \mathrm{Poisson}(\lambda_{t,h}),
\qquad 
\lambda_{t,h} = \mathrm{softplus}\bigl(\mathrm{MLP}_\lambda(\mathbf{h}_{t,h})\bigr).
\]

\paragraph{Negative Binomial (NB) likelihood (over-dispersed counts).}
When counts exhibit over-dispersion but not zero inflation,
\[
Y_{t+h} \sim \mathrm{NB}(\mu_{t,h},\,\theta),
\]
with global dispersion $\theta$ shared across horizons, analogous to the ZINB 
head but without the zero-inflation parameter.

\paragraph{Zero-Inflated Negative Binomial (ZINB) likelihood (sparse, bursty counts).}
Used in our application:
\[
Y_{t+h} \sim \mathrm{ZINB}(\mu_{t,h},\,\pi_{t,h},\,\theta),
\]
requiring the head to output
\[
\mu_{t,h},\qquad \pi_{t,h},\qquad \theta.
\]
The ZINB head adds a dedicated MLP branch for the zero-inflation logits and is 
the most expressive option for zero-heavy conflict data.

Thus, the only components that must be rewritten to adopt a new likelihood 
are the final parametric head and its associated negative log-likelihood. All other parts of the 
architecture: gates, encoder, decoder, rolling optimization, remain unchanged. Table~\ref{tab:likelihood_components} summarizes the facts.
This modular design of DynAttn implements a \emph{general} forecasting framework, not tied to a single distributional assumption. Extending the model to a new 
likelihoods, like general L\'evy or stable distributions, is possible but not-necessarily trivial as this modification requires appropriate reparameterizations, the availability of
numerically stable negative log-likelihood implementations, and careful calibration across horizons.

\begin{table}[h!]
\centering
\caption{Components of DynAttnn affected by the choice of likelihood.}
\label{tab:likelihood_components}
{\small
\begin{tabular}{p{5cm}c p{6cm}}
\toprule
\textbf{Component} & \textbf{Affected?} & \textbf{Explanation} \\
\midrule
Windowing \& standardization & No & Independent of outcome family. \\
Softplus feature gates & No & Gates applied before temporal encoding. \\
Self-attention encoder & No & Learns $\mathbf{h}_t$ regardless of likelihood. \\
Multi-horizon decoder & No & Produces $\mathbf{h}_{t,h}$ for all likelihoods. \\
\textbf{Likelihood head} & \textbf{Yes} & Outputs parameters of the chosen distribution. \\
Loss function (negative log-likelihood) & \textbf{Yes} & Distribution-specific log-likelihood. \\
Exceedance probabilities & Partial & Defined only for distributions with survival functions (Poisson, NB, ZINB, $t$, etc.). \\
\bottomrule
\end{tabular}
}
\end{table}

\section{Data}\label{sec:data}

Our empirical analysis is based on the data infrastructure developed for the ``Fast forward: Forecasting global emerging threats'' competion \citep{DND2025FastForward}, which provides a standardized, high-dimensional
spatio-temporal dataset for forecasting conflict-related fatalities at both the
country and subnational grid levels. The dataset integrates harmonized outcome
variables derived from the Uppsala Conflict Data Program with a broad set of
time-varying and structural covariates covering demographic, economic,
environmental, and accessibility dimensions. All data sources, preprocessing
steps, and variable definitions follow the official competition specifications
and documentation.

The prediction target is the monthly number of conflict-related fatalities as
coded by the Uppsala Conflict Data Program (UCDP), covering state-based violence,
non-state violence, and one-sided violence. Event-level data are aggregated to
monthly counts for each spatial unit. The temporal coverage spans 1990--2025 (July),
providing a long historical window characterized by extreme sparsity, excess
zeros, and heavy-tailed distributions of fatalities.

We conduct the analysis at two spatial resolutions. At the country level, the
dataset includes 191 countries with monthly observations. At the subnational
level, outcomes are defined on the PRIO-Grid, a global lattice of approximately
$0.5^\circ \times 0.5^\circ$ grid cells (roughly $50 \times 50$ km at the equator),
yielding 13{,}110 grid cells with at least one non-missing observation during the
sample period. Grid-level outcomes are constructed by spatially assigning UCDP
events to PRIO-Grid cells and aggregating fatalities within each cell-month.

\subsection{Covariates}

The predictor set comprises several hundred covariates organized into thematic
groups. These include high-frequency conflict-history variables (lagged fatalities,
decay terms, time-since-last-event indicators, and spatial spillovers), climate and
environmental indicators (such as drought severity measures, SPEI-based indices,
and growing-season anomalies), demographic and socioeconomic variables (population,
economic activity proxies, infant mortality), land-use and accessibility measures,
and geographic distance indicators.

All predictors are lagged appropriately to ensure that forecasts rely exclusively
on information available at time $t$, thereby preventing any look-ahead bias.
Country-level and grid-level models draw on the same conceptual set of covariates,
with variables aggregated or spatially disaggregated as required by the unit of
analysis.

\subsection{Benchmark forecasts}

In addition to model-based forecasts generated within DynAttn and other competing
architectures, we include predictions from the VIEWS \cite{Hegre2019VIEWS, Hegre2021VIEWS2020} forecasting system (VIEWS) as an
external benchmark. VIEWS forecast are available via VIEWS API \citep{views2025datasets}. We consider the VIEWS forecast as an additional competing model.

\section{Evaluation of Model Performance}\label{sec:perf}
For both the country and grid level analysis, the data are available till July 2025. We use the period August 2024 –July 2025 (12 months) as test set in our model performance analysis, while the rest of the data is used by each model to train their respective forecasting engines. We include past VIEWS forecasts aligned with the training set so that VIEWS forecasts are considered as a separate model, the target benchmark model.

\subsection{The Validation Data}
\begin{figure}[H]
      \centering    \includegraphics[width=0.75\linewidth]{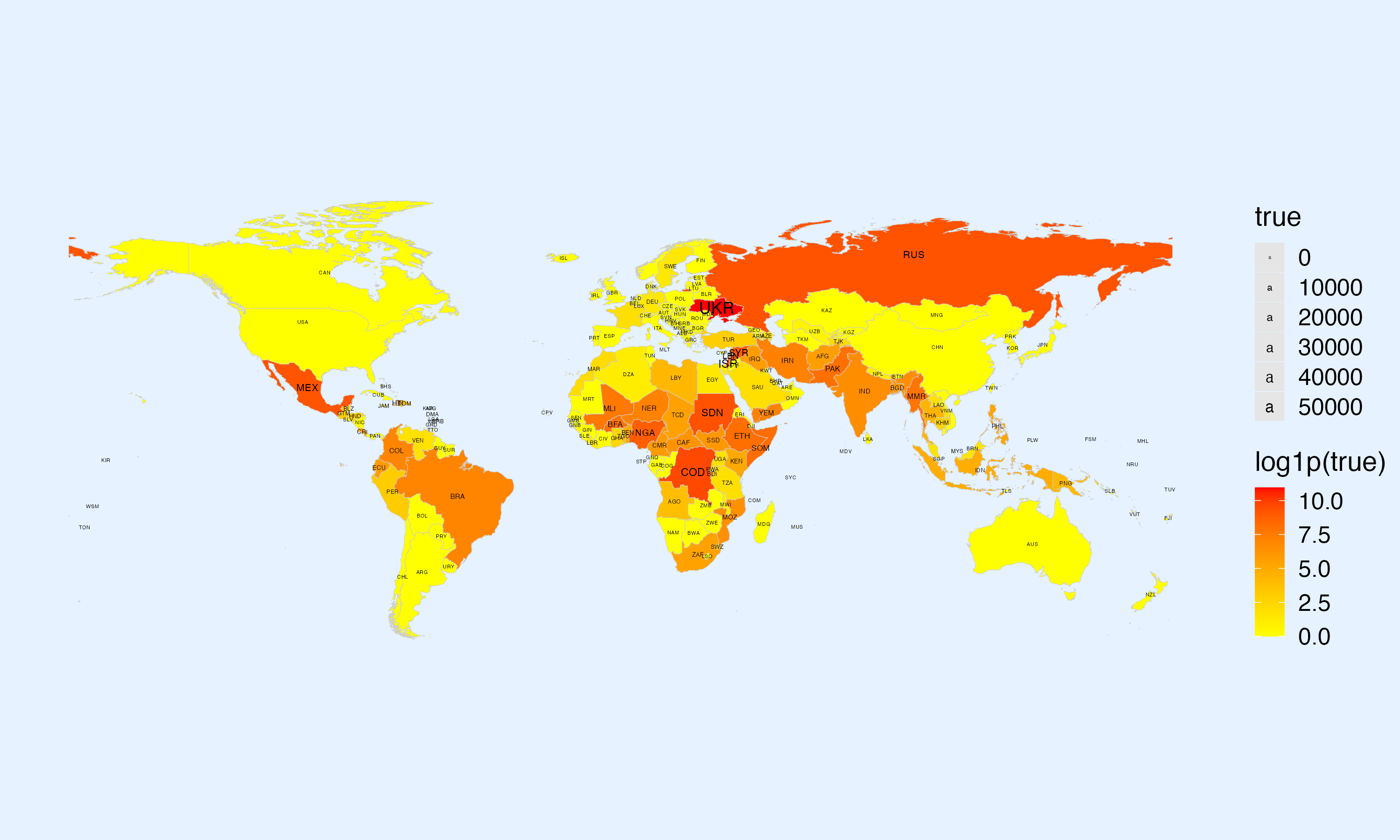}
    \caption{Observed number of fatalities by country in testing period, August 2024 --July 2025. Darker colors indicate more conflict fatalities transformed by $\log(Y+1)$, and larger country labels indicate greater absolute number of fatalities. Given PRIO's country definition, the territory of Israel also contains Palestine. }
    \label{fig:map_true_country}
\end{figure}

\begin{figure}[H]
    \centering
\includegraphics[width=0.75\linewidth]{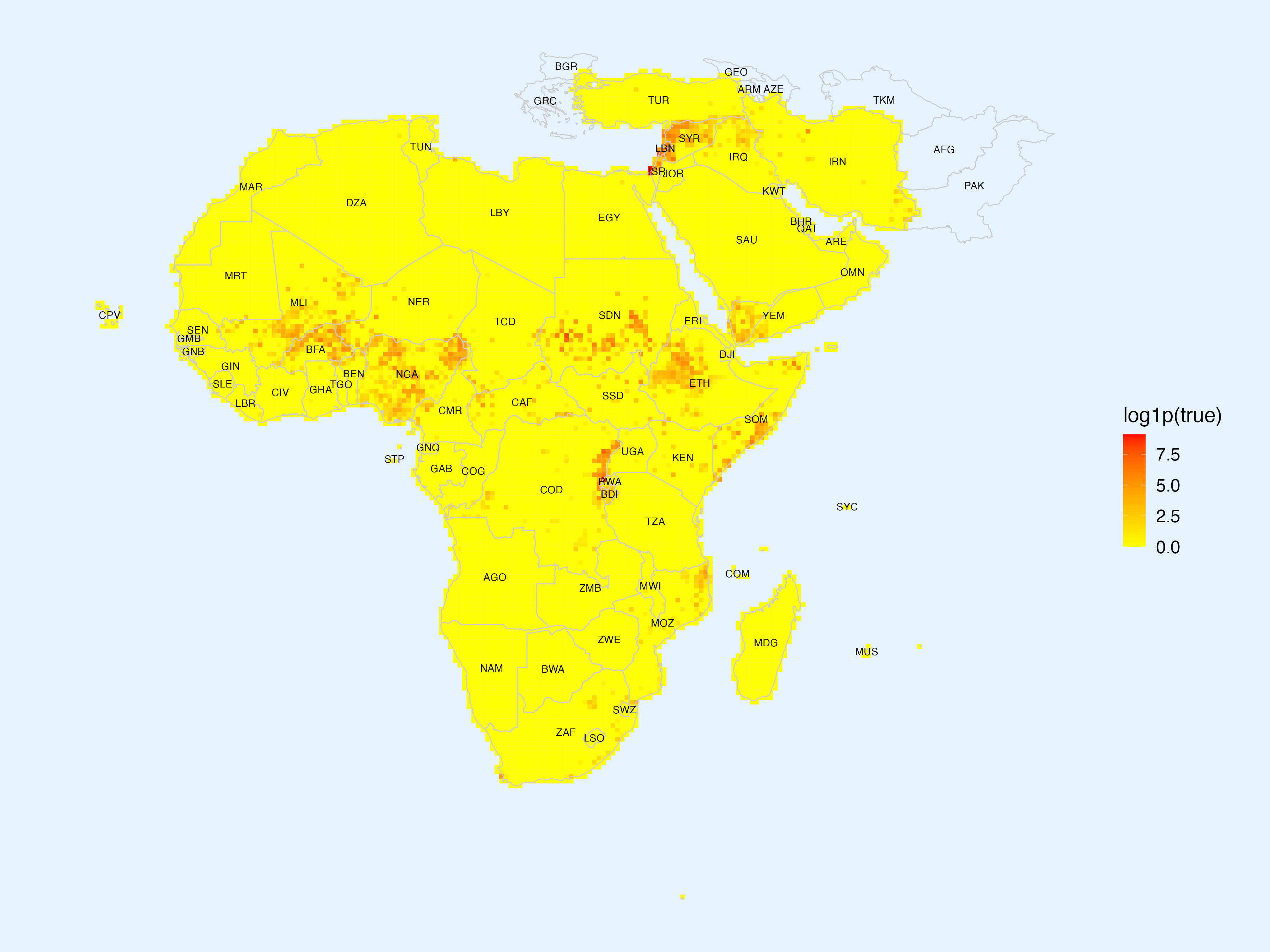}
    \caption{Observed number of fatalities by grid in testing set, August 2024 --July 2025. Darker colors indicate more conflict fatalities transformed by $\log(Y+1)$.}
        \label{fig:map_true_grid}
\end{figure}

Figure~\ref{fig:map_true_country} illustrates the observed number of deaths around the globe during the 12-month testing/validation period (August 2024 --July 2025). Over this testing period, the world was witnessing a high prevalence of conflict; 77 countries (or 40\%) has one or more conflict-related death(s). Among those countries experienced conflict, the human costs do vary; the fatality is the highest in Ukraine (57,166), followed by Israel including Palestine (17,427), Democratic Republic of the Congo (14,537), Sudan (10,506), Russia (10,243), and Mexico (10,029). 
Similarly, Figure~\ref{fig:map_true_grid} depicts the observed fatality at grid-level for the entire Africa and part of middle east. There are 13,110 grids in total, of which 1168 or 8.9\% have at least one fatality.  The highest number of deaths occurred in a grid bordering Democratic Republic of the Congo and Rwanda (8,814 casualties), followed by two grids in Israel including Palestine with deaths 7,296 and 6,926.

\subsection{Forecasting Causalities}
\begin{figure}[]
     \centering
    \includegraphics[width=1\linewidth]{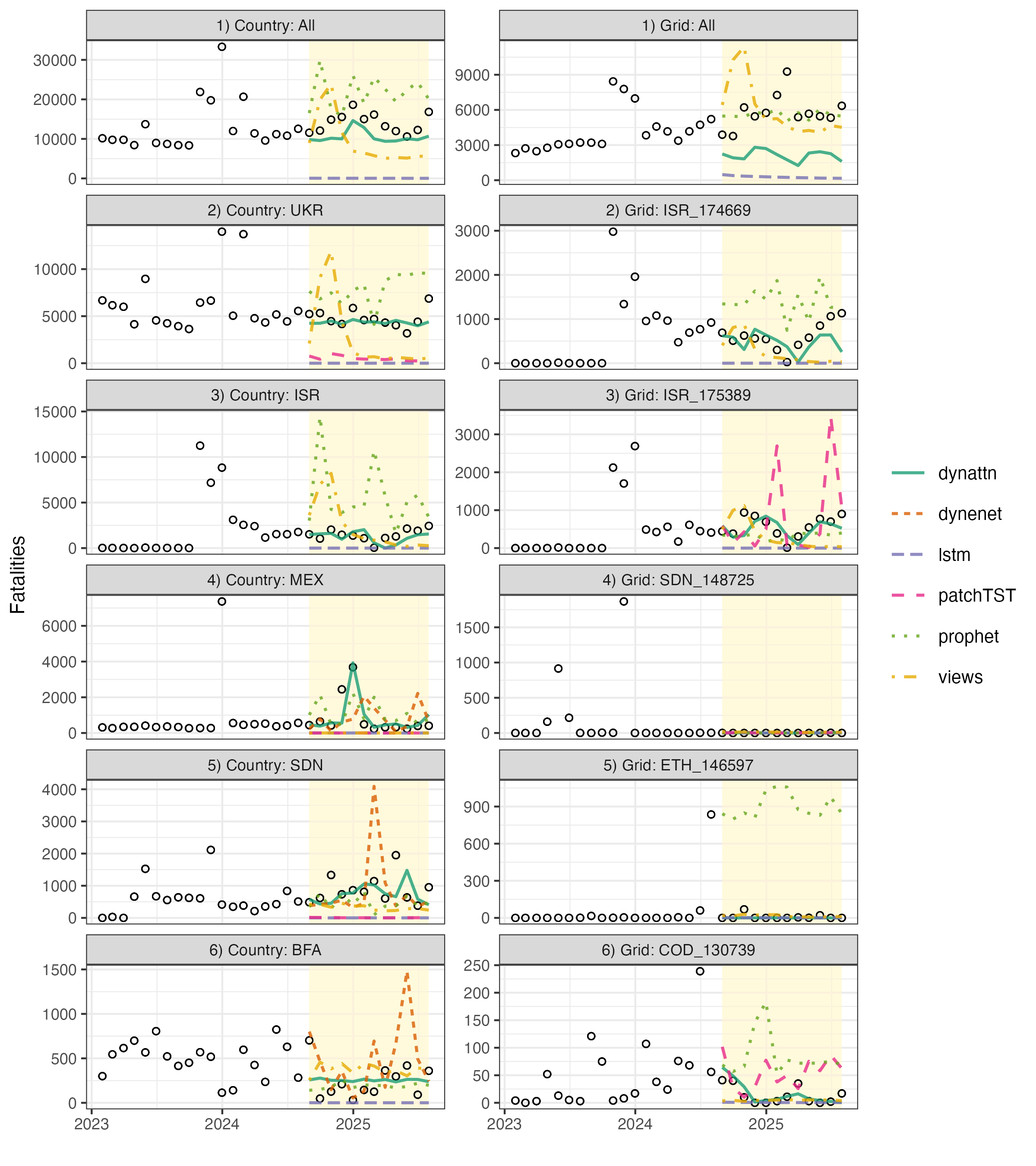}
    \caption{Observed and predicted trends of fatalities. Observed values in circles and predicted values in lines.}
    \label{fig:pred_trend}
\end{figure}

Figure~\ref{fig:pred_trend}  presents the observed and predicted number of fatalities over time for selected countries (left panels) and grids (right panels). The selected countries and grids represent the geographical units with the highest fatalities during the period January 2023 -- July 2024.  The shaded region indicates the testing period, which comprises the last twelve time points used to evaluate the predictive accuracy of DynAttn and benchmark its performance against state-of-the-art models, including DynENet, LSTM, Prophet, PatchTST, and VIEWS. The top two panels show the aggregate fatalities across all countries or all grids. The subsequent panels correspond to the geographic units with the highest number of fatalities during the training period, providing insight into model performance in regions with most intense conflict dynamics.

For the country-level predictions, DynAttn demonstrates strong alignment with observed trends across diverse contexts such as Ukraine (UKR), Israel (ISR), and Mexico (MEX). Competing models exhibit pronounced variances and/or biases in their predictions during the testing window, whereas DynAttn maintains stable and accurate forecasts even in the presence of abrupt changes (e.g., MEX). This robustness underscores DynAttn’s ability to capture temporal dependencies and adapt to sudden shifts in conflict intensity.

For the grid-level predictions, the forecasting challenge is amplified by sparse and highly localized fatality events. Despite this complexity, DynAttn consistently outperforms alternative approaches, delivering smooth and realistic predictions for small areas in Israel, whereby conflict situations were most dire. In contrast, PatchTST, Prophet, and VIEWS often produce exaggerated peaks and erratic fluctuations, highlighting their limitations in handling conflict forecasting in small areas whereby data is highly sparse. LSTM tends to systematically underestimate the true counts of fatalities, and DynENet often fails to produce predictions as its training process were incomplete for many grids.

\subsubsection{Correlation Analysis}
\begin{figure}[]
       \centering
    \includegraphics[scale=0.6]{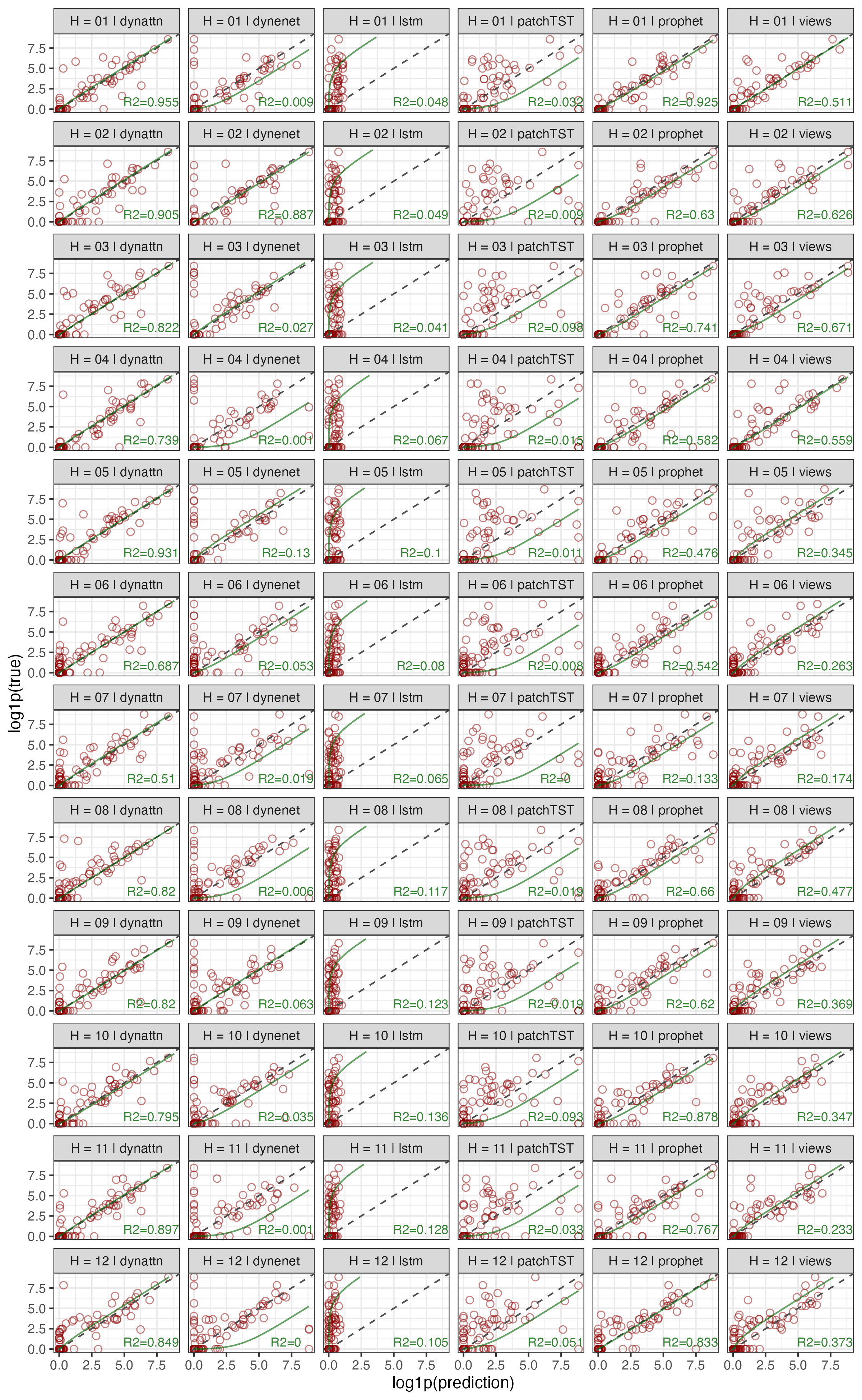}
     \caption{Models' goodness of fit for country-level predictions. The panels show scatter plots of predicted versus observed fatalities for all data points in the testing set, with axes on a log scale ($\log(1+Y)$ transformation applied for visualization purposes only). The green curves represent regression fits, and the reported $R^2$ values are computed on the original (untransformed) scale of fatalities.}
    \label{fig:r2_cm}
\end{figure}

\begin{figure}[H]
    \centering
    \includegraphics[scale=0.6]{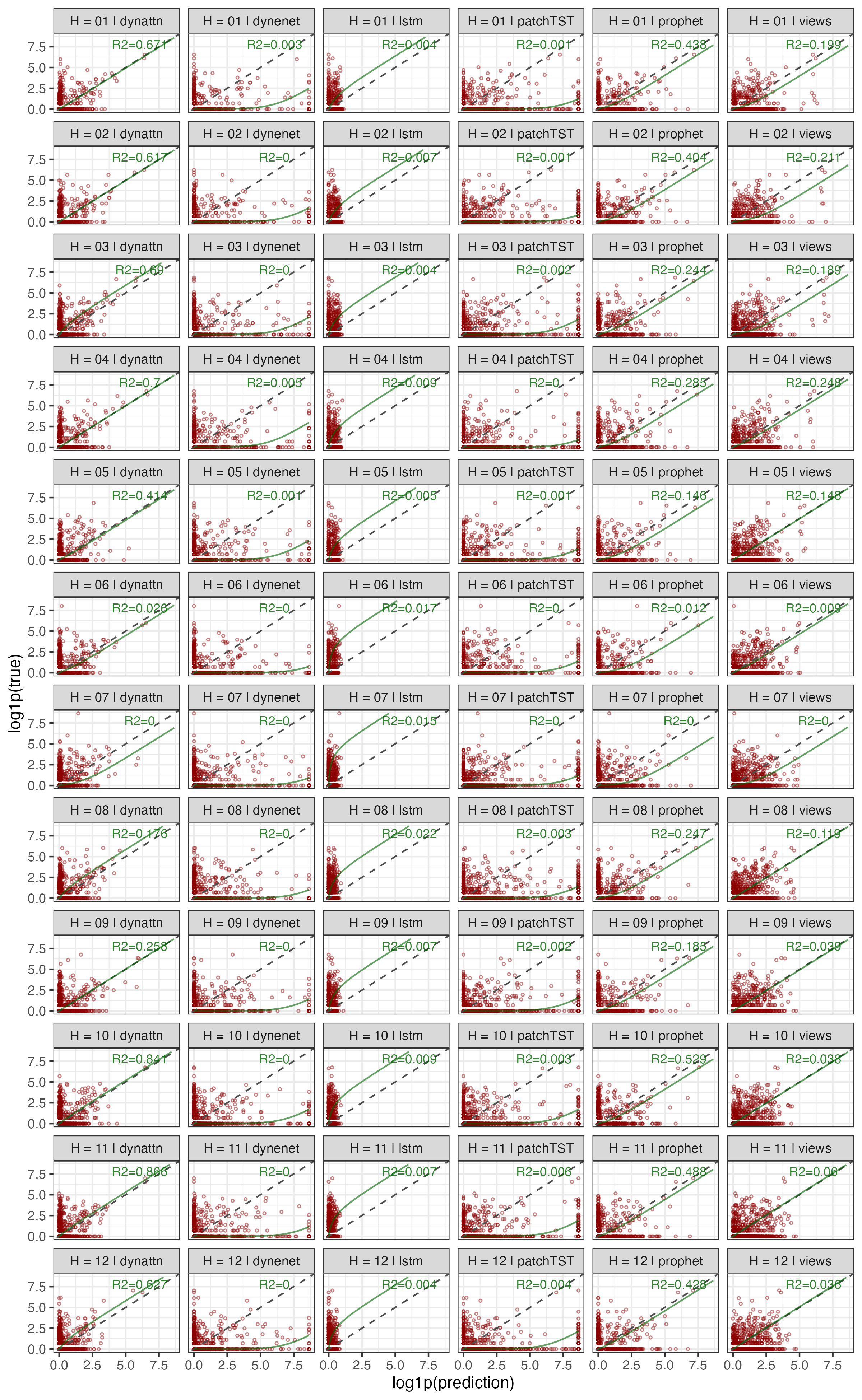}
    \caption{Models' goodness of fit for grid-level predictions. The panels show scatter plots of predicted versus observed fatalities for all data points in the testing set, with axes on a log scale ($\log(1+Y)$ transformation applied for visualization). The green curves represent regression fits, and the reported $R^2$ values are computed on the original (untransformed) scale of fatalities.}
    \label{fig:r2_pgm}
\end{figure}

To assess the predictive capability of our proposed DynAttn algorithm, we compared its performance against four established time-series forecasting methods: Dyenet, Long Short-Term Memory (LSTM), PatchTST, and Prophet—along with a baseline forecast by Views. Figure~\ref{fig:r2_cm} and Figure~\ref{fig:r2_pgm} summarize the predictive performance of DynAttn and competing models using the coefficient of determination ($R^2$) as the evaluation metric for the alignment between the true conflict fatality counts and the predicted values. The scatter plots compare predicted versus observed fatalities across 12 forecasting horizons (row-wise) and 6 models (column-wise). The results unequivocally demonstrate the substantial superiority of the DynAttn model across all tested forecasting horizons. 

\begin{center}
\textbf{[Figure~\ref{fig:r2_cm} about here]}
\end{center}

For the country-level task, DynAttn consistently achieved the highest $R^2$ values from the shortest horizon ($H=01$, $R^2=0.955$) through the longest horizon ($H=12$, $R^2=0.897$). This sustained high performance indicates that the model effectively captures the underlying dynamics of conflict severity, even for long-range predictions. The scatter points for DynAttn exhibit a remarkably tight clustering along the diagonal line of perfect prediction. This visual fidelity confirms the high accuracy and low variance of DynAttn's forecasts. In stark contrast, all competing models—Dyenet, LSTM, PatchTST, and Prophet—showed poor predictive power, particularly for longer horizons. Their $R^2$ scores frequently dropped below $0.2$ and some even approaching zero, indicating that their predictions had negligible correlation with the true fatality counts.

For the grid-level task (Figure~\ref{fig:r2_pgm}), the forecasting challenge is more difficult due to sparse and localized fatality patterns. Nevertheless, DynAttn substantially outperforms all other models with an average $R^2$  of 0.576 across 12 horizons, while Prophet and VIEWS achieve 0.248 and 0.169, respectively. DynENet, LSTM, and PatchTST again perform poorly, with an average $R^2$  values close to zero. Temporal patterns confirm DynAttn’s superior performance across horizons, particularly in later periods where competing models exhibit sharp declines or near-zero predictive power, whereas DynAttn maintains its relatively high level of accuracy.

\begin{center}
\textbf{[Figure~\ref{fig:r2_pgm} about here]}
\end{center}

\subsubsection{Mean Squared Error Analysis}
In addition to correlation-based metrics, we report the root mean squared error (RMSE)
of the predicted fatality counts for each forecast horizon $h = 1, \ldots, 12$ and
each model $m$:
\begin{equation}
\mathrm{RMSE}_m(h)
=
\sqrt{
\frac{1}{N}
\sum_{i=1}^{N}
\left(
\hat{y}_{i,h} - y_{i,h}
\right)^2
},
\end{equation}
where $\hat{y}_{i,h}$ and $y_{i,h}$ denote the predicted and observed fatality counts,
respectively, for spatial unit $i$ at horizon $h$, and $N$ is the number of units.
The index $i$ runs over all countries or PRIO grid cells, depending on the level
of analysis.

\begin{landscape}
    
\begin{table}[ht]
\centering
\begin{adjustbox}{
    max width=\linewidth, max height=\textheight, center}
\tiny
\begin{tabular}{lccccccccccccc}
  \toprule
model  & na.val & H01 & H02 & H03 & H04 & H05 & H06 & H07 & H08 & H09 & H10 & H11 & H12 \\ 
  \midrule
DynAttn &   zero & {\bf 104.97}   & {\bf 139.95}   & {\bf 174.64}   & {\bf 217.59}   & {\bf 149.99}   & {\bf 249.33}   & {\bf 403.66}   & {\bf 152.20}   & {\bf 149.65}   & {\bf 153.92}   & {\bf  123.60}   & {\bf 278.28}   \\ 
  DynENet &   zero & 438.36   & 289.32   & 406.58   & 1e+08   & 508.88   & 453.33   & 6e+53   & 2,719.29   & 336.24   & 310.54   & 2e+12   & 8e+27   \\ 
  LSTM &   zero & 405.63   & 415.27   & 414.61   & 423.69   & 541.92   & 451.63   & 573.19   & 362.58   & 350.21   & 298.29   & 373.22   & 568.79   \\ 
  PatchTST &   zero & Inf   & Inf   & 3e+13   & Inf   & 9e+23   & Inf   & 1e+10   & 2e+29   & Inf   & 1e+28   & Inf   & 1e+14   \\ 
  Prophet &   zero & 238.70   & 994.39   & 346.69   & 353.72   & 581.79   & 465.38   & 900.96   & 512.92   & 511.99   & 520.95   & 523.70   & 304.95   \\ 
  VIEWS &   zero & 281.65   & 525.72   & 725.91   & 287.96   & 461.80   & 396.06   & 531.46   & 298.01   & 293.91   & 247.84   & 330.09   & 504.81  \\
  \midrule
  DynAttn &   drop & {\bf 104.97}   & {\bf 139.95}   & {\bf 174.64}   & {\bf 217.59}   & {\bf 149.99}   & {\bf 249.33}   & {\bf 403.66}   & {\bf 152.20}   & {\bf 149.65}   & {\bf 153.92}   & {\bf  123.60}   & {\bf 278.28}   \\
  DynENet &   drop & 256.14 (107) & 376.01 (105) & 104.16 (106) & 1e+08 (105) & 324.67 (103) & 389.22 (104) & 7e+53 (109) & 3,525.14 (112) & 176.14 (110) & 182.02 (109) & 3e+12 (110) & 1e+28 (110) \\ 
  LSTM &   drop & 405.63   & 415.27   & 414.61   & 423.69   & 541.92   & 451.63   & 573.19   & 362.58   & 350.21   & 298.29   & 373.22   & 568.79   \\ 
  PatchTST &   drop & Inf   & Inf   & 3e+13   & Inf   & 9e+23   & Inf   & 1e+10   & 2e+29   & Inf   & 1e+28   & Inf   & 1e+14   \\ 
  Prophet &   drop & 239.96 (189) & 999.64 (189) & 348.52 (189) & 355.58 (189) & 584.86 (189) & 467.84 (189) & 905.72 (189) & 515.63 (189) & 514.70 (189) & 523.70 (189) & 526.46 (189) & 306.56 (189) \\ 
  VIEWS &   drop & 281.65   & 525.72   & 725.91   & 287.96   & 461.80   & 396.06   & 531.46   & 298.01   & 293.91   & 247.84   & 330.09   & 504.81   \\
   \bottomrule
\end{tabular}
\end{adjustbox}
\caption{Model-specific predictive errors (RMSE) for each forecast horizon averaged over all countries. Top and bottom panels include forecasts with and without missing values, respectively. Values in parentheses indicate the number of valid predictions. If not specified, model trainings are completed and hence RMSE are calculated, for all countries (N=191). In bold the lowest RMSE's.}
\label{tab:RMSE_country}
\end{table}

\begin{table}[ht]
\centering
\begin{adjustbox}{
    max width=\linewidth, max height=\textheight, center}
\tiny
\begin{tabular}{llllllllllllll}
  \toprule
model  & na.val & H01 & H02 & H03 & H04 & H05 & H06 & H07 & H08 & H09 & H10 & H11 & H12 \\ 
  \midrule
DynAttn &   zero & {\bf 4.99}   & {\bf 4.46}   & {\bf 9.71}   &{\bf  6.05}   &{\bf 9.36}   & {\bf 26.97}   & 50.11   & 7.25   & {\bf 10.03}   & {\bf 4.68}   &{\bf  5.21}   & {\bf 10.96}   \\ 
  DynENet &   zero & Inf   & Inf   & Inf   & Inf   & Inf   & Inf   & Inf   & 4e+132   & Inf   & Inf   & Inf   & Inf   \\ 
  LSTM &   zero & 8.65   & 7.18   & 12.93   & 11.03   & 11.99   & 27.13   & 49.97   & 7.71   & 11.64   & 10.92   & 11.72   & 14.56   \\ 
  PatchTST &   zero & Inf   & Inf   & Inf   & Inf   & Inf   & Inf   & Inf   & Inf   & Inf   & Inf   & Inf   & Inf   \\ 
  Prophet &   zero & 11.25   & 11.48   & 14.00   & 14.44   & 16.13   & 31.62   & 51.31   & 14.40   & 12.82   & 13.72   & 10.83   & 12.07   \\ 
  VIEWS &   zero & 10.43   & 18.68   & 22.16   & 10.15   & 11.07   & 27.04   & {\bf 50.05}   & {\bf 7.24}   & 11.41   & 10.71   & 11.36   & 14.30  \\ 
  \midrule
 DynAttn &   drop & {\bf 4.99}   & {\bf 4.46}   & {\bf 9.71}   &{\bf  6.05}   &{\bf 9.36}   & {\bf 26.97}   & 50.11   & 7.25   & {\bf 10.03}   & {\bf 4.68}   &{\bf  5.21}   & {\bf 10.96}   \\ 
  DynENet &   drop & Inf (3287) & Inf (3340) & Inf (3352) & Inf (3384) & Inf (3374) & Inf (3382) & Inf (3406) & 8e+132 (3436) & Inf (3454) & Inf (3455) & Inf (3457) & Inf (3469) \\ 
  LSTM &   drop & 8.65   & 7.18   & 12.93   & 11.03   & 11.99   & 27.13   & 49.97   & 7.71   & 11.64   & 10.92   & 11.72   & 14.56   \\ 
  PatchTST &   drop & Inf   & Inf   & Inf   & Inf   & Inf   & Inf   & Inf   & Inf   & Inf   & Inf   & Inf   & Inf   \\ 
  Prophet &   drop & 11.25   & 11.48   & 14.00   & 14.44   & 16.13   & 31.62   & 51.31   & 14.40   & 12.82   & 13.72   & 10.83   & 12.07   \\ 
  VIEWS &   drop & 10.43   & 18.68   & 22.16   & 10.15   & 11.07   & 27.04   & {\bf 50.05}   & {\bf 7.24}   & {\bf 11.41}   & 10.71   & 11.36   & 14.30   \\  
   \bottomrule
\end{tabular}
\end{adjustbox}
\caption{Model-specific predictive errors (RMSE) for each forecast horizon averaged over all grids. Top and bottom panels include forecasts with and without missing values, respectively. Values in parentheses indicate the number of valid predictions. If not specified, model trainings are completed and hence RMSE are calculated for all grids (N=13110). In bold the lowest RMSE's.}
\label{tab:RMSE_grid}
\end{table}

\end{landscape}

\subsection{Predicting the Risk of Conflict with DynAtt}
DynAttn not only predicts expected fatality counts, but also estimates the probability
that violence exceeds a specified threshold $\tau \ge 0$. For each forecast horizon
$h$, the model produces exceedance probabilities of the form
\begin{equation}
\Pr\!\left( Y_{t+h} \ge \tau \,\middle|\, \mathcal{F}_t \right),
\end{equation}
where $Y_{t+h}$ denotes the number of conflict-related fatalities at horizon $t+h$,
$\tau$ is a predefined intensity threshold, and $\mathcal{F}_t$ represents the
information set available at time $t$. These probabilities are obtained analytically
from the ZINB likelihood head, which explicitly models zero inflation, the conditional
mean, and overdispersion, thereby enabling exceedance-based inference rather than
only point predictions.

\begin{figure}[H]
    \centering
\includegraphics[width=1\linewidth]{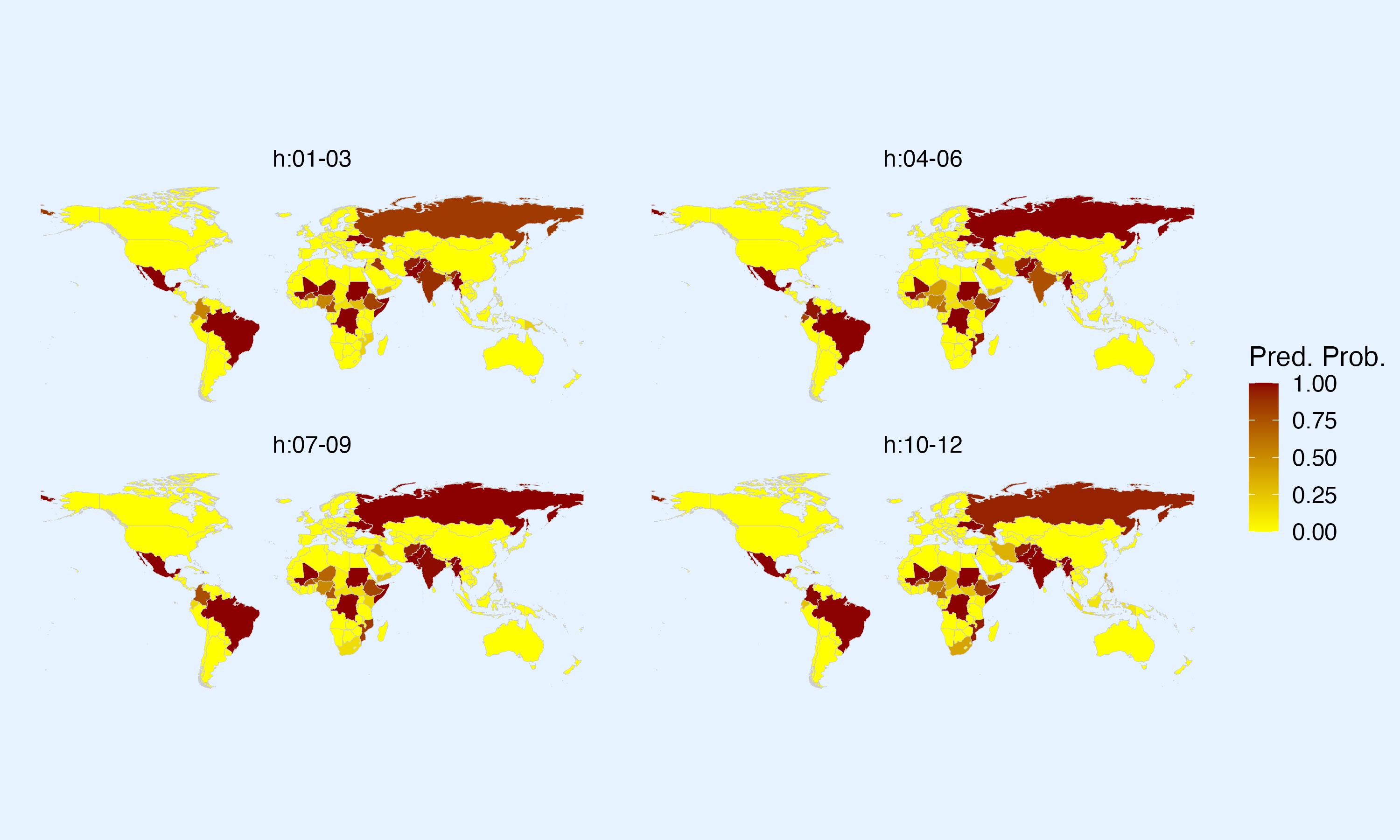}
    \caption{Predicted risk of at least 25 fatalities in testing set, August 2024 --July 2025. Darker colors indicate higher probability of encountering conflict fatalities.}
        \label{fig:cm_map_dynattn_predprob}
\end{figure}

\begin{figure}[H]
    \centering
\includegraphics[width=1\linewidth]{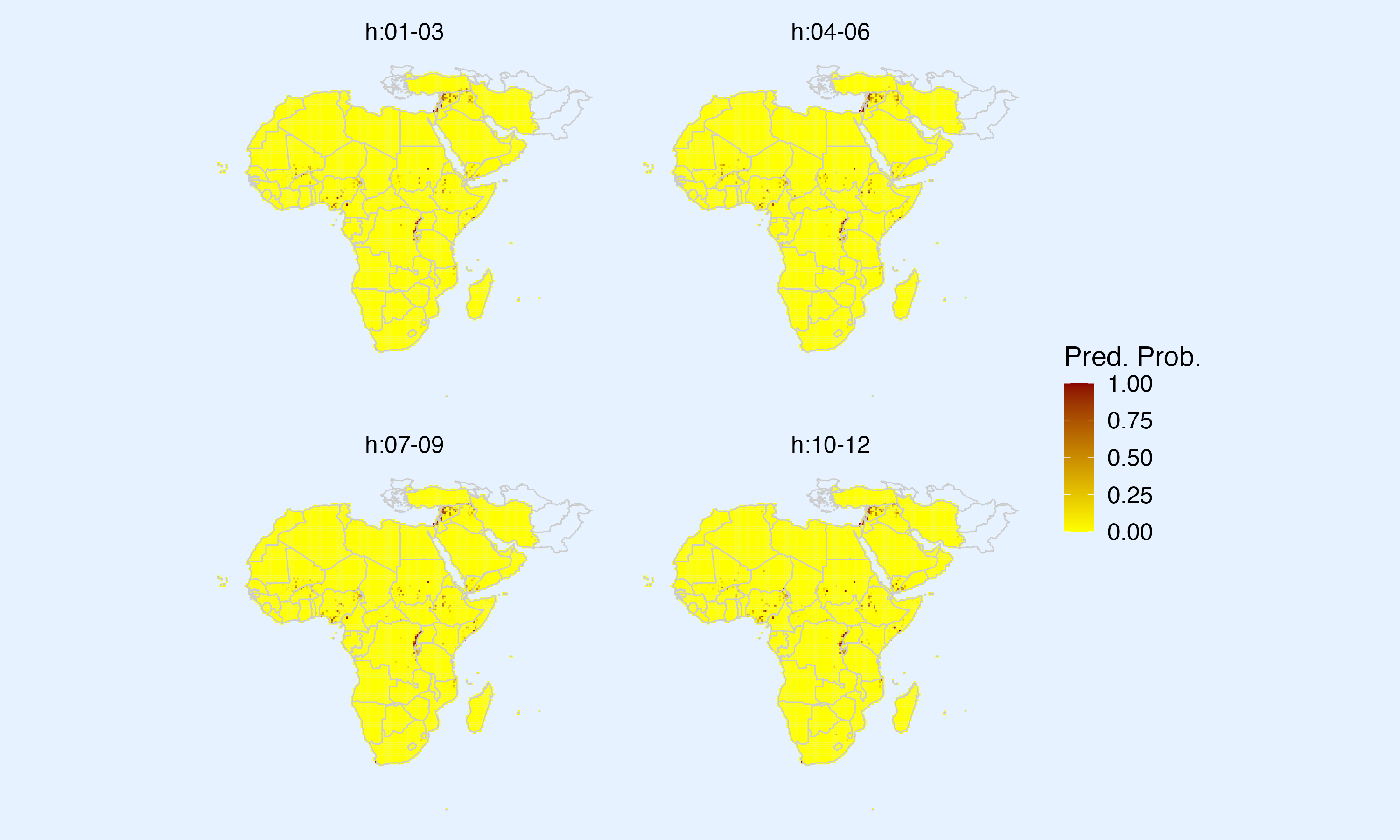}
    \caption{Predicted risk of at least 1 fatalities in testing set, August 2024 --July 2025. Darker colors indicate higher probability of encountering conflict fatalities.}
        \label{fig:pgm_map_dynattn_predprob}
\end{figure}

The resulting exceedance probabilities admit a natural interpretation in terms of
conflict risk, such as the likelihood of moderate- or high-intensity violence,
depending on the chosen threshold $\tau$. Figures~\ref{fig:cm_map_dynattn_predprob}
and~\ref{fig:pgm_map_dynattn_predprob} display quarterly-averaged risk maps for
country-level ($\tau \ge 25$) and grid-level ($\tau \ge 1$) predictions, respectively,
aggregated over horizons H1--3, H4--6, H7--9, and H10--12.

\subsection{Summary of the Performance Analysis}

In summary, the rigorous cross-model comparisons above suggest that the DynAttn
architecture brings a significant advancement for the quantitative analysis of
conflict. As the key findings demonstrate, DynAttn's predictions consistently exhibit
a close alignment with observed fatalities, reflected by its sustained high coefficient
of determination ($R^2$) across nearly all forecasting horizons. Conversely, alternative
deep learning and statistical models more frequently fail to capture variance in the
data, often yielding $R^2$ values approaching zero.

These conclusions are further corroborated by error-based evaluation metrics.
Across all forecast horizons, DynAttn systematically achieves the lowest root mean
squared error (RMSE) among competing models, both at the country and grid levels
(Tables~\ref{tab:RMSE_country} and~\ref{tab:RMSE_grid}). The stability of these RMSE
gains across short-, medium-, and longer-term horizons indicates that DynAttn not
only explains a larger share of variance, but also produces more accurate point
forecasts in absolute terms.

Taken together, these results underscore DynAttn's efficacy in addressing the unique
methodological challenges posed by sparse and highly volatile conflict time series.
They further reveal the model's superior ability to capture complex and non-linear
spatiotemporal dynamics and to adapt to abrupt changes in conflict intensity. This
capability provides political scientists and policymakers with a more reliable and
actionable tool for forecasting armed conflict.

\section{Explainability of the Forecasts}

A unique feature of DynAttn, compared to other deep learning algorithms, is its explainability. Specifically, the model has a feature gating mechanism, effectively reducing the dimensionality of the covariate space. Not only does this mechanism ensure model parsimony, and thus yielding stable predictions, it also allows us to further explore the complex relations between the gated features and target variable through ablation and elasticity. Below, we demonstrate how DynAttn's explanability is operationalized.

Recall that, in DynAttn, each feature $j=1,\dots,F$, is modulated by a nonnegative gate,
\[
g_j \;=\; \max\!\big\{0, \mathrm{softplus}(\gamma_j) - \varepsilon\big\}, \qquad \varepsilon=10^{-4}.
\]
We then rank the magnitude of these nonnegative gates and retain the top fraction,
\[
\mathcal{J}_{\rho} \;=\; \big\{ j \,:\, g_j \text{ in top $\rho$ quantile of $\{g_1,\dots,g_F\}$} \big\},
\]
Here, we use a threshold $\rho=0.10$ (i.e., top 10\% of positively gated features). These gated features indexed by $  j \in \mathcal{J}_{\rho}$ are then mapped into DynAttn's downstream diagnostics, quantifying their importance through ablation and elasticity calculations.

\subsection{Ablation importance}
Ablation is a procedure for evaluating a given feature's importance by comparing model's predictive performance (e.g., prediction errors) with and without it. For each gated feature $j \in \mathcal{J}_{\rho}$ at each evaluation anchor $\tau$, the baseline expected count is:
\[
\widehat{y}_\tau \;=\; \mathbb{E}[Y_\tau \mid \mathbf{X}_\tau] \;=\; (1-\pi_\tau)\mu_\tau
\]
The squared error is:
\[
\mathrm{SE}_\tau = \big(\widehat{y}_\tau - y_\tau\big)^2.
\]
To ablate, we force feature $j$'s gate to zero, yielding a counterfactual  $\widehat{y}^{(-j)}_\tau$ and compute its corresponding squared error $\mathrm{SE}^{(-j)}_\tau$. The ablation importance of feature $j$ is the average relative change in MSE:
\[
\Delta_j \;=\; \frac{1}{|\mathcal{T}|} \sum_{\tau \in \mathcal{T}}
   \frac{\mathrm{SE}^{(-j)}_\tau - \mathrm{SE}_\tau}{\mathrm{SE}_\tau},
\]
where $\mathcal{T}$ indexes the evaluation anchors. A large positive $\Delta_j$ indicates that feature $j$
is critical for predictive accuracy. This measure explains the robustness of the model to the predictors.

\subsection{Elasticity}
Elasticity measures the percentage change in the outcome with respect to a percentage change in a given feature. For each feature $j \in \mathcal{J}_{\rho}$ at evaluation anchor $\tau$, and consider the last standardized
window $\mathbf{X}_\tau$ and its baseline expected count $\widehat{y}_\tau$, we form a perturbed window $\mathbf{X}_\tau^{(+\delta,j)}$ by applying a $+\delta$ shock to feature $j$:
\[
\mathbf{x}_{t,j}^{(+\delta)} \;=\; (1+\delta) \, \mathbf{x}_{t,j}, 
\qquad t=\tau-S+1,\dots,\tau,
\]
leaving all other coordinates unchanged. Passing this window through the model yields $\widehat{y}^{(+\delta,j)}_\tau$. The elasticity of feature $j$ is defined as the average percent effect:
\[
\mathrm{Elas}_j \;=\; \frac{1}{|\mathcal{T}|} \sum_{\tau \in \mathcal{T}}
   \frac{\widehat{y}^{(+\delta,j)}_\tau - \widehat{y}_\tau}{|\widehat{y}_\tau| + \epsilon},
\]
with $\delta=0.10$ (a $+10\%$ shock) and a small $\epsilon$ for numerical stability. Positive/negative values indicate the percentage increase/decrease in the expected casualties with respect to a 10\% increase/decrease in  feature $j$.

The workflow elaborated above reports (i) the selected set $\mathcal{J}_{0.10}$, (ii) ablation importance $\Delta_j$, and (iii) elasticities $\mathrm{Elas}_j$ for each selected feature. These diagnostics provide an interpretable quantification of how
predictors influence forecasts, complementing raw gate magnitudes.

Figure~\ref{fig:cm_tile}  illustrate the gate values for each feature $j$, as well as the ablation and elasticity for feature $  j \in \mathcal{J}_{\rho}$ at country-level. Each figure contains three panels corresponding to (i) gate magnitudes, (ii) elasticity, and (iii) ablation importance for the selected features. The ordering of both axes in Figure~\ref{fig:cm_tile}  reflects Ward.D hierarchical clustering, which groups countries and features according to similarity in their gating. For visualization purpose, we normalized the ablation and elasticity as:
\begin{align*}
            & \kappa_{i,j}  = \dfrac{1}{12} \sum^{12}_{h=1} \kappa_{i,j,h}  \\
            & \kappa'_{i,j} = 
\begin{cases} 
\dfrac{\kappa_{i,j}}{\max_{j} \left | \kappa_{i,j} \right |},  & \text{if }  \max_{j} \left | \kappa_{i,j} \right | > 0 \\
0, & \text{if } \max_{j} \left | \kappa_{i,j} \right | = 0 
\end{cases}
\end{align*}
where, $\kappa$  denotes $\Delta$ or $\mathrm{Elas}$ . $i, j,h$ denotes geographical unit, feature, and horizon, respectively.

\section{Data Driven Explanation of Conflicts}\label{sec:understand}

While DynAttn is designed primarily for forecasting, its structure also enables a
systematic interpretation of which predictors are most informative for short- to
medium-horizon conflict dynamics. Because conflict data are high-dimensional, sparse,
and strongly zero-inflated, naive importance measures can be unstable or misleading.
We therefore rely on three complementary diagnostics that are naturally supported by
the DynAttn architecture. First, the shared elastic-net gating step performs
regularized feature selection, yielding a sparse set of predictors with non-zero gate
mass and thereby isolating the variables that the model consistently relies on across
units and horizons. Second, elasticity measures quantify the sensitivity of the
forecast to marginal changes in each selected predictor, providing a scale-free
summary of effect strength that supports cross-variable comparison. Third, ablation
analysis evaluates how predictive performance degrades when a feature (or feature
group) is removed, offering an outcome-oriented notion of importance based on
forecast error rather than on parameter magnitude alone. Together, these tools
provide an interpretable map from a high-dimensional predictor space to empirically
salient drivers of forecasted fatalities, while preserving the distinction between
predictive relevance and causal interpretation. We apply this framework first at the
country level and then at high spatial resolution using PRIO grid cells.

\subsection{Country Level Analysis}
\begin{figure}[H]
    \centering
    \includegraphics[width=1\linewidth]{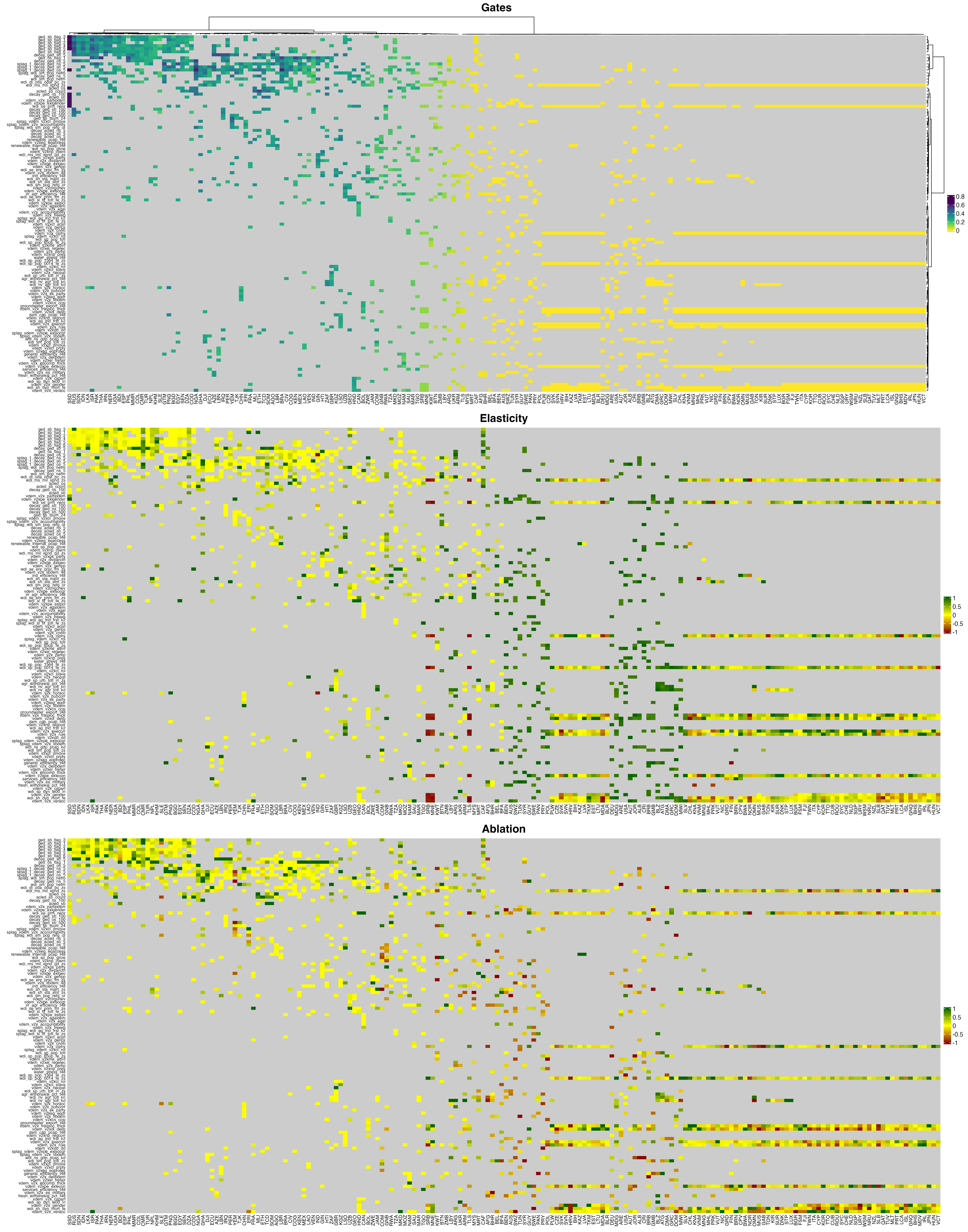}
     \caption{Country-level feature gates, elasticities, and ablations.}
    \label{fig:cm_tile}
\end{figure}

Across countries, we observe three main clusters in gating. The top-left cluster in  Figure~\ref{fig:cm_tile} suggests that countries (from SSD to COD) tend to select various time lags of state-based conflict fatalities into $\mathcal{J}_{0.10}$ with strong magnitude. In the second cluster (from NGA to ARM), the gated features remain strong in magnitude, but become diverse. The third large cluster (from TLS to VCT) consists of countries whereby the selected features are weak in magnitude with gate values close to zero, suggesting that the given set of predictors is not predictive for fatalities in these countries. It is noteworthy that countries with most intense conflict, such as RUS, SDN, ISR, belongs to the top left cluster, in which the number of fatalities is strongly correlated with its own time lags, indicating persistency of conflict as an overall predictor. The second cluster is also notable as it includes UKR, MEX, RWA which are also the hotspots of conflict. However, the factors driving conflicts appear to be much more heterogeneous, compared to the first cluster.

Although time lags of fatalities appear to be important gates in the first cluster, their magnitudes of elasticity are unevenly distributed (see second panel of Figure~\ref{fig:cm_tile}); there are only few normalized elasticities close to 1, suggesting that fatalities in each country is associated with a specific lag length, and the lag length do vary across countries, revealing heterogeneous temporality. Similarly, in the second cluster, despite diverse features being selected, their magnitudes are concentrated in certain variables. For example, in UKR, fatality vary most strongly with \texttt{wdi\_sp\_dyn\_le00\_in} (life expectancy at birth), whereas, in MEX and RWA, it is most strongly associated with \texttt{ind\_efficiency\_t48} (Industrial Water Use Efficiency). These are structural features of the countries in which conflict arise.
\begin{center}
\textbf{[Figure~\ref{fig:cm_tile} about here]}
\end{center}
In terms of ablation, the patterns tend to be more complex, compared to elasticity. Some countries tend to have multiple features influencing model's predictive performance. For example, in RWA, the ablation importance is above 0.4 for features: \texttt{decay\_ged\_sb\_100}, \texttt{decay\_ged\_sb\_500}, \texttt{ged\_sb\_tlag\_5}, and \texttt{wdi\_sh\_sta\_stnt\_zs}, suggesting that if the model does not account for the lags or decay of the target variable, or excludes the prevalence of stunting among children, the MSE of predictions would have increased by 40\%.

\subsection{High Resolution Spatial Analysis of Conflicts}
\begin{figure}[H]    
    \centering
    \includegraphics[width=1\linewidth]{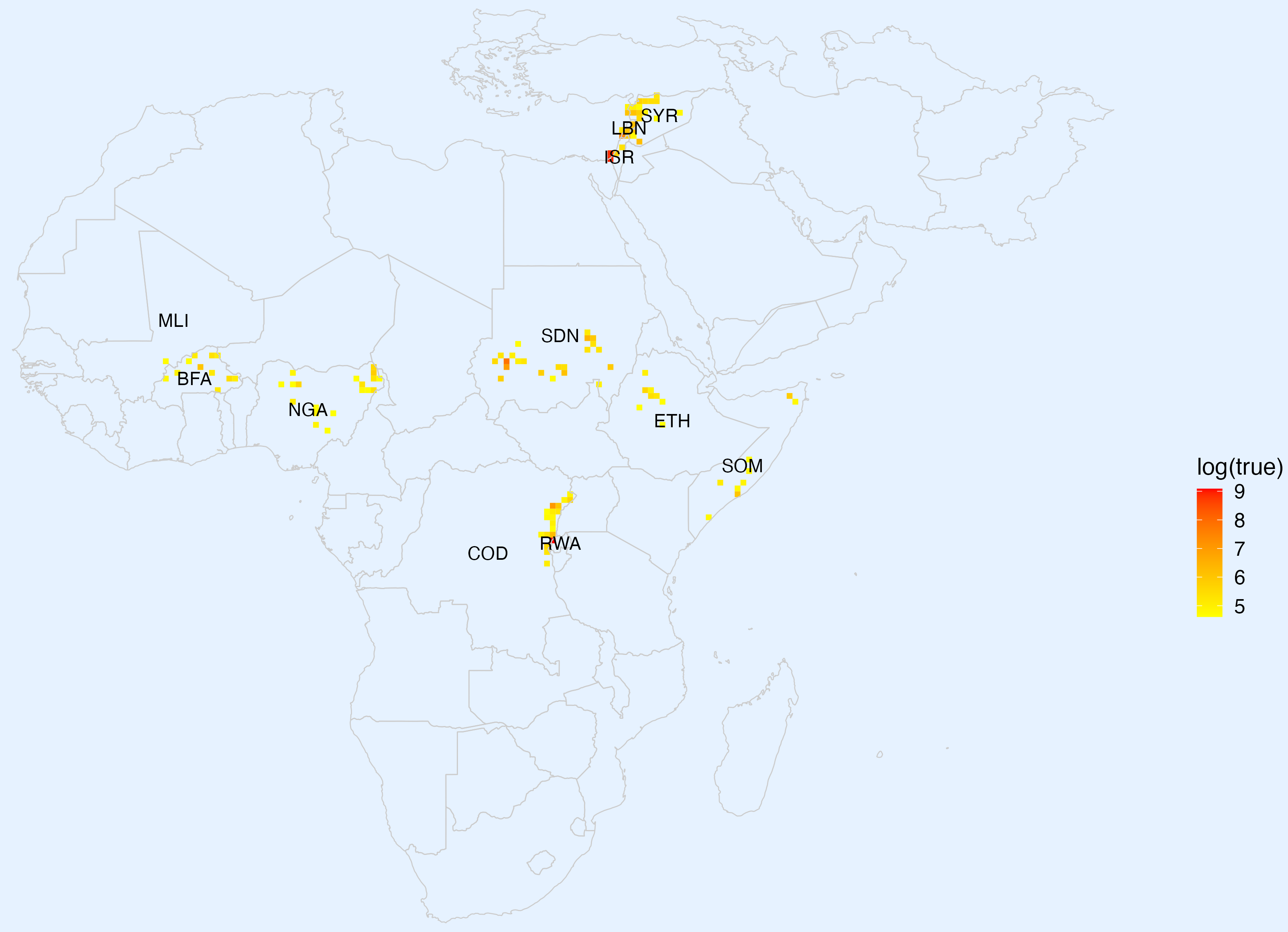}
    \caption{Conflict hotspots - grids with fatalities $>100$}
    \label{fig:selgrid}
\end{figure}

The grid-level model allows us to uncover factors underlying conflict fatalities with high spatial resolution. Here, we select four cases where inter-state and/or internal conflicts tend to be most intense: Western Africa (NGA-BFA-MLI), Eastern Africa (SDN-ETH-SOM), Central Africa (COD-RWA-BDI), and Middle East (ISR-LBN-SYR). As shown in Figure~\ref{fig:selgrid}, these cases constitute the hotspots whereby the number of fatalities exceeds 100 in each grid. 

For each conflict we extract the relevant explanatory variables that have been used to generate the forecasts and highlight similarities and differences.

\begin{center}
\textbf{[Figure~\ref{fig:selgrid} about here]}
\end{center}

\subsubsection{Central/East Africa: COD--RWA--BDI}
The COD-RAW-BDI system is not so much a war between DRC, Rwanda and Burundi in conventional, officially declared war sense,  but rather a complex, still-ongoing conflict in eastern DRC involving a rebel group  \textit{March 23 Movement} (often called M23),  widely considered backed by Rwanda. The conflict is part of a broader set of tensions in eastern DRC, especially in the Kivu region, dating back decades.
While the highest casualty concentrations are clearly localized along the eastern
DRC--Rwanda border, Burundi enters the COD--RWA--BDI system primarily through
cross-border spillovers and regional contagion rather than sustained high-fatality
violence on its own territory.
In June 2025, DRC and Rwanda signed a peace accord (brokered by the U.S. and Qatar), aiming to end Rwandan support for M23 and get a ceasefire. 
Yet, despite the deal, fighting has continued. Rwandan troops have not fully withdrawn and M23 has kept operating (and advancing) in eastern DRC.

\begin{figure}[H]
    \centering
    \includegraphics[width=1\linewidth]{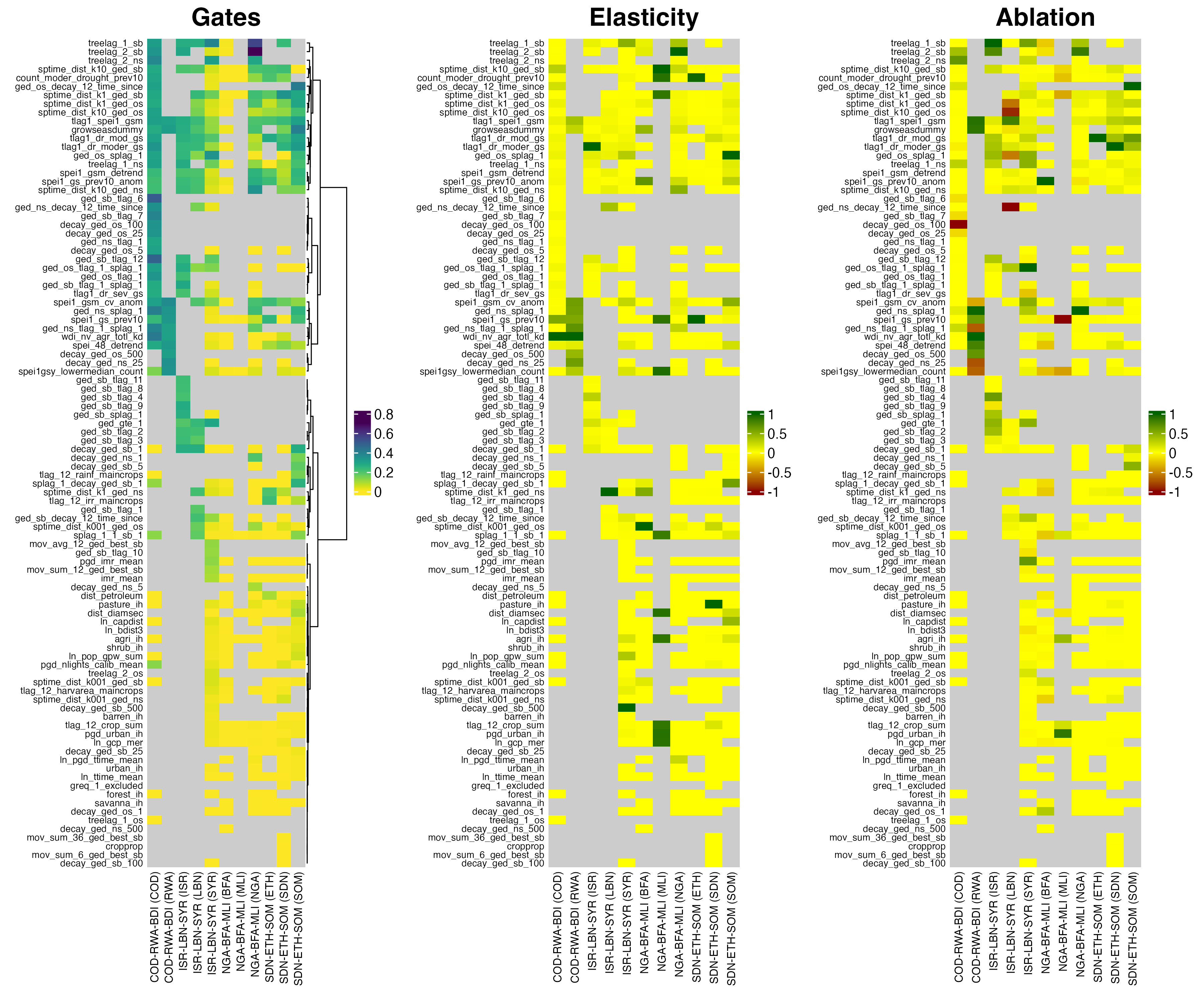}
     \caption{PRIO grid-level feature gates, elasticities, and ablations average by conflict region.}
    \label{fig:pgm_tile}
\end{figure}

What does our model says about the predictors on the number of casualties in this region?
For the COD–RWA-BDI system, violence is strongly path-dependent and spatially contagious.
The model learns that recent conflict in nearby locations is the single most informative
signal for future violence, consistent with well-documented mechanisms such as
rebel mobility across porous borders, retaliatory dynamics, and localized escalation
cycles in eastern DRC.
This aligns with the empirical literature on self-exciting conflict processes, but here
these dynamics are learned directly from the data through the DynAttn architecture,
rather than being imposed ex ante. Table~\ref{tab:gate_COD_RWA_BDI} reports the average value of the gates parameters from the elastic net step of the DynAttn model while Figure~\ref{fig:cm_tile} show country specific gate values and Figure~\ref{fig:pgm_tile} shows average ablation and elasticity measures at conflict level.
\begin{center}
\textbf{[Table~\ref{tab:gate_COD_RWA_BDI} about here]}
\end{center}
\begin{center}
\textbf{[Figure~\ref{fig:pgm_tile} about here]}
\end{center}
\paragraph{Conflict persistence and local diffusion.}
The model assigns high relevance to predictors capturing temporal persistence,
spatial diffusion, event-type specificity, and recency effects in patterns of
violence. Temporal persistence is reflected in lagged measures at 1, 6, and
12 months, including \texttt{ged\_sb\_tlag\_6}, \texttt{ged\_sb\_tlag\_12},
\texttt{ged\_ns\_splag\_1}, \texttt{ged\_ns\_tlag\_1}, and
\texttt{ged\_os\_tlag\_1}. Spatial diffusion is captured through spillovers from
nearby grid cells, such as \texttt{treelag\_1\_sb}, \texttt{treelag\_2\_sb}, and
\texttt{treelag\_2\_ns}. Event-type specificity is encoded via joint lag–spillover
terms distinguishing state-based, non-state, and one-sided violence
(\texttt{ged\_*\_tlag\_*\_splag\_*}, \texttt{decay\_ged\_*}), while recency effects
are represented by decay measures and time-since-last-event indicators
(\texttt{*\_time\_since}).

All these predictors receive gate values in the range $[0.35, 0.50]$ for the
COD–RWA regional conflict system.

\paragraph{Regional conflict entanglement (cross-border structure).}
The model also assigns substantial gate mass to variables capturing
spatio-temporal proximity to violence beyond purely local history.
These include spatial–temporal distance measures and spillover terms such as
\texttt{sptime\_dist\_k1\_*}, \texttt{sptime\_dist\_k10\_*}, \texttt{splag\_*},
as well as joint lag–spillover interactions.

These predictors encode exposure to nearby violence in both space and time,
rather than isolated within-cell dynamics.
Substantively, the high relevance of these variables indicates that the model
places considerable weight on violence occurring in geographically proximate
locations and recent periods, reflecting the cross-border and regionalized
structure of the COD–RWA-BDI conflict system.
This is consistent with mechanisms such as armed group mobility across borders,
displacement-driven spillovers, and escalation along transport routes and border
corridors.
Taken together, these patterns support treating COD–RWA-BDI as a single regional
conflict theater rather than as independent national conflict processes.

\paragraph{Climate stress as a conditional amplifier.}
Climate-related predictors receive moderate gate values (approximately
$[0.20, 0.32]$), including
\texttt{spei1\_gsm}, \texttt{spei1\_gsm\_cv\_anom}, \texttt{spei\_48\_detrend},
\texttt{growseasdummy}, \texttt{count\_moder\_drought\_prev10}, and drought
severity indicators such as \texttt{dr\_moder\_gs} and \texttt{dr\_sev\_gs}.

These variables contribute to the forecast primarily after conditioning on
recent conflict dynamics.
Substantively, this indicates that environmental stress operates as a conflict
multiplier rather than a primary driver: droughts and rainfall anomalies increase
conflict volatility, but mainly in contexts where organized violence is already
present.
This pattern is consistent with expectations for eastern DRC, where climate
shocks tend to amplify competition and vulnerability in areas with existing armed
capacity.
Importantly, this hierarchy emerges endogenously through the gating mechanism,
rather than being imposed by model specification.

\paragraph{Weak selection of structural economic variables.}
In contrast, classic slow-moving structural covariates receive low gate values
(approximately $[0.00, 0.13]$). These include population size
(\texttt{ln\_pop\_gpw\_sum}), economic activity proxies
(\texttt{pgd\_nlights\_calib\_mean}), land-use indicators
(\texttt{forest\_ih}, \texttt{agri\_ih}, \texttt{pasture\_ih}),
distance to petroleum infrastructure (\texttt{dist\_petroleum}),
and capital distance (\texttt{ln\_capdist}).

Substantively, this indicates that for short- to medium-horizon forecasting in
the COD–RWA system, baseline development, population, and geographic
structure add little incremental predictive power once recent violence and
diffusion dynamics are accounted for.
This does not imply that such factors are unimportant determinants of conflict
risk in a structural sense; rather, they are comparatively weak short-run
predictors relative to dynamic, high-frequency signals.
Making this distinction explicit is important to avoid conflating predictive
relevance with causal importance.

\subsubsection{Middle East: ISR--LBN--SYR}
The ISR--LBN--SYR system corresponds to a highly interconnected regional conflict
environment shaped by recurrent cross-border hostilities, asymmetric warfare, and
overlapping state and non-state actors. While Israel, Lebanon, and Syria differ
substantially in terms of institutional capacity and conflict intensity, violence in
this region is tightly coupled across borders through recurrent escalation episodes,
localized spillovers, and persistent low-intensity confrontations.

What does our model say about the predictors of casualties in this region?
For the ISR--LBN--SYR system, the forecasting signal is again dominated by endogenous
conflict dynamics, but with a more heterogeneous temporal structure than, for example, in Central
Africa. The model assigns substantial relevance to short- and medium-run lags of
state-based and non-state violence, as well as to spatial spillovers, indicating that
casualty dynamics are shaped by repeated escalation cycles rather than by purely
long-memory persistence. These patterns are consistent with episodic flare-ups,
cross-border exchanges of fire, and retaliatory dynamics characterizing the
Israel--Lebanon--Syria theater.
Table~\ref{tab:gate_ISR-LBN-SYR} reports the average gate values from the elastic-net
selection step of the DynAttn model, while Figure~\ref{fig:selgrid} illustrates the
corresponding spatial distribution of selected grid cells.
\begin{center}
\textbf{[Table~\ref{tab:gate_ISR-LBN-SYR} about here]}
\end{center}

\paragraph{Conflict persistence and episodic escalation.}
The model assigns high relevance to lagged measures of recent violence, particularly
non-state and state-based events, though with a broad dispersion across time lags. Prominent predictors include
\texttt{treelag\_2\_ns}, \texttt{treelag\_1\_sb},
\texttt{ged\_sb\_tlag\_9}, \texttt{ged\_sb\_tlag\_12},
\texttt{ged\_sb\_tlag\_11}, and \texttt{ged\_sb\_tlag\_2}.
This pattern suggests that violence in the region is not only persistent, but also
characterized by repeated episodic escalation over several months, rather than by
simple short-run autoregression.
Event-type specificity further matters, as reflected by the relevance of
\texttt{ged\_os\_tlag\_1}, \texttt{ged\_os\_splag\_1}, and joint lag–spillover terms.
Recency effects, captured through decay and time-since-last-event indicators such as
\texttt{ged\_sb\_decay\_12\_time\_since}, remain informative but receive lower gate
values than in the Central African system. Overall, the highest gate values for this
group lie in the range $[0.20, 0.32]$.

\paragraph{Regional conflict entanglement and cross-border spillovers.}
Spatial and spatio-temporal proximity to violence plays a central role in the
ISR--LBN--SYR system. The model assigns substantial gate mass to spillover and distance
measures, including \texttt{splag\_*}, \texttt{sptime\_dist\_k1\_ged\_sb}, and
\texttt{sptime\_dist\_k10\_ged\_ns}, as well as multiple
joint lag–spillover interactions.
These predictors capture exposure to violence occurring in geographically proximate
areas and recent periods, rather than isolated local dynamics.

Substantively, this reflects the highly interconnected nature of the regional conflict
theater, where violence frequently propagates across borders through airstrikes,
rocket fire, militia activity, and localized retaliatory actions. The model therefore
supports treating ISR--LBN--SYR as a tightly coupled regional system, rather than as a
set of independent national conflict processes.

\paragraph{Climate stress and contextual modifiers.}
Climate-related predictors receive moderate but non-negligible gate values in the
ISR--LBN--SYR system. Variables such as \texttt{tlag1\_spei1\_gsm},
\texttt{spei1\_gsm\_cv\_anom}, \texttt{spei1\_gsm\_detrend}, and
\texttt{growseasdummy} contribute to the forecast primarily after conditioning on
recent conflict dynamics.

Substantively, this indicates that environmental stress acts as a contextual modifier
rather than a primary driver of casualties. Climatic variability may influence
operational conditions, population vulnerability, or timing of escalation, but its
predictive relevance remains secondary to endogenous conflict dynamics in this
highly militarized environment.

\paragraph{Limited role of structural and socioeconomic covariates.}
As in other regions, slow-moving structural covariates receive comparatively low gate
values. Population size (\texttt{ln\_pop\_gpw\_sum}), economic activity proxies
(\texttt{pgd\_nlights\_calib\_mean}, \texttt{ln\_gcp\_mer}), land-use indicators, and
geographic distance measures (\texttt{ln\_capdist}, \texttt{ln\_bdist3}) add little
incremental predictive power once recent violence and spillover dynamics are accounted
for.
Some demographic and health-related proxies, such as infant mortality
(\texttt{pgd\_imr\_mean}, \texttt{imr\_mean}), enter with modest gate values, suggesting
a limited role for underlying vulnerability in shaping casualty risk at short to
medium forecasting horizons. As before, this does not imply that these factors are
unimportant in a structural sense, but rather that they are weak short-run predictors
relative to dynamic, high-frequency signals of conflict.

\subsubsection{West Africa (Sahel): NGA--BFA--MLI}
The NGA--BFA--MLI system corresponds to the central Sahelian conflict theater,
characterized by highly mobile non-state armed groups, weak territorial control,
and recurrent cross-border violence across Nigeria, Burkina Faso, and Mali.
Unlike more centralized interstate or proxy conflicts, violence in this region is
predominantly driven by decentralized insurgent activity, fluid frontlines, and
strong interactions between environmental stress and local security conditions.

What does our model say about the predictors of casualties in this region?
For the NGA--BFA--MLI system, the DynAttn model identifies a combination of
short-run conflict dynamics and climate-related stressors as the dominant
predictive signals. Compared to the Central/East Africa and Middle East systems,
environmental variables play a substantially more prominent role, while long-run
conflict persistence is weaker. Table~\ref{tab:gate_NGA-BFA-MLI} reports the average
gate values from the elastic-net selection step of the DynAttn model.
\begin{center}
\textbf{[Table~\ref{tab:gate_NGA-BFA-MLI} about here]}
\end{center}

\paragraph{Conflict persistence and non-state violence dynamics.}
The model assigns its highest gate values to predictors capturing recent
non-state violence and its diffusion across nearby locations. Prominent variables
include \texttt{treelag\_2\_ns}, \texttt{ged\_ns\_splag\_1},
\texttt{ged\_ns\_tlag\_1\_splag\_1}, and decay measures such as
\texttt{decay\_ged\_ns\_1} and \texttt{decay\_ged\_ns\_5}.
State-based violence enters the model primarily through spillover and tree-lag
terms (e.g., \texttt{treelag\_1\_sb}, \texttt{treelag\_2\_sb}), rather than through
long autoregressive lags.

This pattern suggests that casualty dynamics in the Sahel are shaped less by
long-memory persistence and more by short-lived escalation episodes associated
with mobile non-state actors and rapid spatial diffusion. The highest gate values
for this group lie approximately in the range $[0.18, 0.29]$.

\paragraph{Climate stress as a primary driver.}
In contrast to the other regions examined, climate-related predictors receive
gate values comparable to, and in some cases exceeding, those of conflict-history
variables. Indicators of short-term and seasonal climatic stress, including
\texttt{spei1\_gsm\_detrend}, \texttt{tlag1\_spei1\_gsm},
\texttt{spei1\_gs\_prev10\_anom}, \texttt{growseasdummy}, and
\texttt{spei1\_gsm\_cv\_anom}, rank among the most influential predictors in the
model.

Substantively, this indicates that environmental stress acts as a first-order
driver of casualty risk in the NGA--BFA--MLI system, rather than merely as a
conditional amplifier. Rainfall deficits, drought anomalies, and seasonal
constraints directly shape mobility, resource competition, and civilian exposure
in contexts where armed groups operate opportunistically and state presence is
limited.

\paragraph{Spatial diffusion and cross-border contagion.}
Spatio-temporal distance and spillover measures also receive non-negligible gate
mass, including \texttt{sptime\_dist\_k1\_ged\_sb},
\texttt{sptime\_dist\_k1\_ged\_os}, \texttt{sptime\_dist\_k10\_ged\_ns}, and related
terms. These variables capture exposure to violence in nearby grid cells and
recent periods, highlighting the porous borders and rapid cross-border movement
of armed actors across Nigeria, Burkina Faso, and Mali.

Together with the prominence of non-state violence indicators, this supports
treating NGA--BFA--MLI as a highly interconnected regional conflict system driven
by diffusion and opportunistic escalation rather than by localized, persistent
conflict equilibria.

\paragraph{Minimal role of structural covariates.}
As in the other regions, slow-moving structural variables receive very low gate
values. Population size, economic activity proxies, land-use indicators, distance
measures, and health-related covariates contribute little incremental predictive
power once dynamic conflict and climate signals are taken into account.

This does not imply that structural conditions are irrelevant for understanding
the long-run emergence of conflict in the Sahel. Rather, it indicates that for
short- to medium-horizon forecasting of casualties, high-frequency indicators of
violence and environmental stress dominate the predictive landscape.

\subsubsection{Horn of Africa: SDN--ETH--SOM}
The SDN--ETH--SOM system corresponds to a highly volatile regional conflict theater
characterized by overlapping civil wars, state fragmentation, and recurrent
humanitarian crises across Sudan, Ethiopia, and Somalia. Violence in this region is
shaped by a combination of one-sided attacks against civilians, fragmented non-state
actors, and strong exposure to environmental stress, with frequent spillovers across
porous borders.

What does our model say about the predictors of casualties in this region?
For the SDN--ETH--SOM system, the DynAttn model identifies a forecasting structure
dominated by recency effects, climate stress, and short-range spatial diffusion.
Compared to the other regions examined, the model assigns exceptionally high relevance
to measures capturing the time since recent violent events, indicating that casualty
risk responds sharply to recent shocks rather than exhibiting long-run persistence.
Table~\ref{tab:gate_SDN_ETH_SOM} reports the average gate values from the elastic-net
selection step of the DynAttn model.
\begin{center}
\textbf{[Table~\ref{tab:gate_SDN_ETH_SOM} about here]}
\end{center}

\paragraph{Recency and one-sided violence dynamics.}
The single most influential predictor in this region is the time since recent
one-sided violence, as captured by
\texttt{ged\_os\_decay\_12\_time\_since}, which receives the highest gate value
($\approx 0.41$). Additional decay and short-memory indicators for both state-based
and non-state violence, such as \texttt{decay\_ged\_sb\_5} and
\texttt{decay\_ged\_ns\_1}, also receive substantial gate mass.

This pattern indicates that casualty dynamics in the Horn of Africa are strongly
driven by abrupt escalation episodes, particularly attacks targeting civilians,
followed by rapid decay rather than by persistent autoregressive conflict processes.
Violence is therefore highly reactive to recent events, consistent with cycles of
mass violence, temporary lulls, and renewed outbreaks observed in the region.

\paragraph{Climate stress as a core driver of violence.}
Climate-related variables receive gate values comparable to, and in some cases
exceeding, those of conflict-history predictors. Indicators of short-term drought and
rainfall stress, including \texttt{tlag1\_dr\_moder\_gs},
\texttt{tlag1\_spei1\_gsm}, \texttt{spei1\_gsm\_detrend},
\texttt{spei1\_gs\_prev10\_anom}, and \texttt{count\_moder\_drought\_prev10}, rank
among the most influential predictors in the model.

Substantively, this suggests that environmental stress acts as a first-order driver of
casualty risk in the SDN--ETH--SOM system. Climatic shocks directly affect livelihoods,
food security, and displacement, increasing civilian exposure to violence in contexts
where state protection is weak and armed actors operate opportunistically. Unlike in
more militarized theaters, climate variability here is not merely a conditional
amplifier but a central determinant of short-run violence dynamics.

\paragraph{Spatial diffusion and regional contagion.}
Spatio-temporal proximity to violence also plays a significant role. The model assigns
non-negligible gate mass to distance and spillover measures such as
\texttt{sptime\_dist\_k1\_ged\_sb}, \texttt{sptime\_dist\_k1\_ged\_os},
\texttt{sptime\_dist\_k10\_ged\_sb}, and \texttt{ged\_ns\_splag\_1}.
These variables capture exposure to violence occurring in nearby grid cells and recent
periods.

Together with the prominence of recency effects, this indicates that violence in the
Horn of Africa propagates through short-range contagion and displacement-driven
spillovers rather than through long-distance or long-memory diffusion processes.

\paragraph{Minimal role of structural covariates.}
As in the other regions, slow-moving structural variables receive very low gate values.
Population size, economic activity proxies, land-use indicators, infrastructure
measures, and health-related covariates add little incremental predictive power once
dynamic conflict and climate signals are taken into account.

This does not imply that structural conditions are irrelevant for understanding the
long-run roots of conflict in the Horn of Africa. Rather, it indicates that for
short- to medium-horizon forecasting of casualties, rapid shifts in violence and
environmental stress dominate the predictive landscape.

\subsection{Cross-regional implications: common structure and regional specificities}
Taken together, the four regional case studies highlight both common regularities and
substantial heterogeneity in the drivers of short- to medium-horizon casualty
dynamics. A first robust pattern is that recent violence (measured through lags, decay
terms, and spatio-temporal proximity) consistently receives non-negligible gate mass
across all regions. This indicates that, regardless of context, high-frequency signals
of conflict history remain essential for forecasting casualties, reflecting persistence,
contagion, and retaliation mechanisms that are well documented in the conflict
literature.

Beyond this shared backbone, however, the model uncovers pronounced region-specific
structures. In the COD--RWA--BDI system, the forecasting signal is dominated by strong
path dependence and spatial diffusion, with climate stress entering primarily as a
conditional amplifier once organized violence is already present. In ISR--LBN--SYR,
endogenous conflict dynamics also dominate, but with a more heterogeneous temporal
signature consistent with episodic escalation cycles, and with climate variables
playing a secondary, contextual role. In the NGA--BFA--MLI Sahelian system, by contrast,
environmental stressors become first-order predictors alongside non-state violence
dynamics, consistent with a setting in which climatic variability directly shapes
mobility, vulnerability, and the opportunity structure for armed actors. Finally, in
SDN--ETH--SOM the model assigns exceptional relevance to recency effects and one-sided
violence dynamics, combined with strong climate sensitivity and short-range diffusion,
capturing a highly reactive conflict environment characterized by abrupt escalations
and rapid decay.

These results underscore a central implication for conflict forecasting: there is no
single universal predictive mechanism that generalizes unchanged across theaters.
Instead, forecasting performance depends on models that can accommodate heterogeneous
mixtures of persistence, diffusion, environmental stress, and vulnerability, with
different components becoming dominant in different regions. DynAttn is designed
precisely for this setting. By combining dynamic attention with feature gating and
region-specific regularization, the model adapts to local predictive structure and
recovers distinctive, empirically interpretable hierarchies of predictors without
hard-coding region-specific assumptions. In this sense, the cross-regional evidence
supports viewing conflict forecasting not as the search for a universal model, but as
the construction of flexible, data-adaptive systems that can represent multiple
forecasting regimes within a single coherent framework.

\section{Conclusions}\label{sec:conclusions}

This paper introduces DynAttn, an interpretable dynamic-attention framework for
forecasting conflict-related fatalities in high-dimensional and sparse
spatio-temporal settings. By integrating rolling-window estimation, shared
elastic-net feature gating, a compact weight-tied self-attention encoder, and a
zero-inflated negative binomial likelihood, DynAttn addresses several long-standing
limitations in conflict forecasting. In particular, it reconciles probabilistic
fidelity for zero-heavy count data with strong predictive performance and
transparent diagnostic tools.

Empirically, DynAttn substantially outperforms a wide range of established
forecasting approaches, including linear dynamic elastic-net models, recurrent
neural networks, Transformer-based time-series architectures, and the operational
VIEWS benchmark. These gains are especially pronounced at the PRIO grid level,
where data sparsity and localization severely challenge conventional models.
Across forecast horizons from one to twelve months, DynAttn maintains high
correlation with observed fatalities while avoiding the instability, attenuation,
or extreme volatility exhibited by competing methods.

Beyond forecasting accuracy, DynAttn enables systematic interpretation of
short- to medium-horizon conflict dynamics. The feature-gating mechanism, combined
with ablation and elasticity analyses, reveals a consistent hierarchy of
predictors across regions. High-frequency signals of recent violence—captured
through lags, decay terms, and spatio-temporal spillovers—form the core predictive
backbone in all conflict systems examined. Climate-related variables play a
context-dependent role, acting as conditional amplifiers in some theaters (such as
Central/East Africa and the Middle East) and as first-order drivers in others (notably
the Sahel and the Horn of Africa). In contrast, slow-moving structural covariates,
including population size, economic activity proxies, and land-use indicators,
provide little incremental predictive power at short forecasting horizons once
dynamic signals are accounted for. Importantly, these findings concern predictive
relevance rather than causal importance, and the two should not be conflated.

The cross-regional evidence underscores a central implication for conflict
forecasting: no single universal mechanism governs short-run casualty dynamics.
Instead, different conflict theaters exhibit distinct mixtures of persistence,
diffusion, environmental stress, and recency effects. DynAttn is explicitly designed
to accommodate this heterogeneity. By learning region-specific feature hierarchies
endogenously, without imposing hard-coded assumptions, the model adapts to diverse
forecasting regimes within a unified and interpretable framework.

Several extensions follow naturally from this work. The modular design of DynAttn
allows alternative likelihoods to be incorporated with minimal architectural
changes, enabling applications beyond conflict fatalities to other sparse and
heavy-tailed outcomes. Future research may also explore joint spatial modeling,
hierarchical pooling across regions, or integration with decision-theoretic
frameworks for early warning and risk prioritization. More broadly, the results
suggest that progress in conflict forecasting depends not only on deeper or larger
models, but on architectures that balance probabilistic correctness,
data-efficiency, and interpretability. DynAttn represents a step in this direction.

\clearpage

\bibliographystyle{plainnat} 
\bibliography{literature}       

@article{Haffner_Hofer_Nagl_Walterskirchen_2023, title={Introducing an Interpretable Deep Learning Approach to Domain-Specific Dictionary Creation: A Use Case for Conflict Prediction}, volume={31}, DOI={10.1017/pan.2023.7}, number={4}, journal={Political Analysis}, author={Häffner, Sonja and Hofer, Martin and Nagl, Maximilian and Walterskirchen, Julian}, year={2023}, pages={481–499}}

@article{CranmerDesmarais17, 
title={What Can We Learn from Predictive Modeling?}, volume={25}, DOI={10.1017/pan.2017.3}, number={2}, journal={Political Analysis}, author={Cranmer, Skyler J. and Desmarais, Bruce A.}, year={2017}, pages={145–166}}

@article{Hegre2021VIEWS2020,
  title   = {ViEWS2020: Revising and evaluating the ViEWS political violence early-warning system},
  author  = {Hegre, H{\aa}vard and Bell, Curtis and Colaresi, Michael and Croicu, Mihai and Hoyles, Frederick and Jansen, Remco and Lindqvist-McGowan, Angelica and Randahl, David and R{\o}d, Espen Geelmuyden and Leis, Maxine Ria and Vesco, Paola},
  journal = {Journal of Peace Research},
  year    = {2021},
  volume  = {58},
  number  = {3},
  pages   = {599--611},
  doi     = {10.1177/0022343320962157}
}

@misc{views2025datasets,
author = "VIEWS",
year = 2024,
  title        = {VIEWS API documentation},
  howpublished = {\url{https://github.com/prio-data/views_api/wiki/Available-datasets}},
  note         = {Accessed: 2025-12-24}
}

@misc{DND2025FastForward,
author={"Government of Canada"},
year = {2024},
  title        = {Forecasting global emerging threats (Fast Forward Emerging Threats competition)},
  howpublished = {\url{https://www.canada.ca/en/department-national-defence/programs/defence-ideas/element/contests/challenge/fast-foward-emerging-threats.html}},
  note         = {Accessed: 2025-12-24}
}

@article{Hegre2019VIEWS,
  title   = {ViEWS: A political violence early-warning system},
  author  = {Hegre, H{\aa}vard and Allansson, Marie and Basedau, Matthias and Colaresi, Michael and Croicu, Mihai and Fjelde, Hanne and Hoyles, Fredrik and Hultman, Lisa and H{\"o}gbladh, Stina and Jansen, Ragnhild and Mouhleb, Nils and Muhammad, Shahzad A. and Nilsson, Desir{\'e}e and Nyg{\aa}rd, H{\aa}vard Mokleiv and Olafsdottir, Gudrun and Petrova, Katerina and Randahl, David and R{\o}d, Espen Geelmuyden and Schneider, Gerald and von Uexkull, Nina and Vestby, J{\"o}ran},
  journal = {Journal of Peace Research},
  year    = {2019},
  volume  = {56},
  number  = {2},
  pages   = {155--174},
  doi     = {10.1177/0022343319823860}
}

@inproceedings{vaswani2017attention,
 author = {Vaswani, Ashish and Shazeer, Noam and Parmar, Niki and Uszkoreit, Jakob and Jones, Llion and Gomez, Aidan N and Kaiser, \L ukasz and Polosukhin, Illia},
 booktitle = {Advances in Neural Information Processing Systems},
 editor = {I. Guyon and U. Von Luxburg and S. Bengio and H. Wallach and R. Fergus and S. Vishwanathan and R. Garnett},
 pages = {},
 publisher = {Curran Associates, Inc.},
 title = {Attention is All you Need},
 url = {https://proceedings.neurips.cc/paper_files/paper/2017/file/3f5ee243547dee91fbd053c1c4a845aa-Paper.pdf},
 volume = {30},
 year = {2017}
}

@inproceedings{zhou2021informer,
  author    = {Haoyi Zhou and
               Shanghang Zhang and
               Jieqi Peng and
               Shuai Zhang and
               Jianxin Li and
               Hui Xiong and
               Wancai Zhang},
  title     = {Informer: Beyond Efficient Transformer for Long Sequence Time-Series Forecasting},
  booktitle = {The Thirty-Fifth {AAAI} Conference on Artificial Intelligence, {AAAI} 2021, Virtual Conference},
  volume    = {35},
  pages     = {11106--11115},
  publisher = {{AAAI} Press},
  year      = {2021},
}

@misc{wu2021autoformer,
      title={Autoformer: Decomposition Transformers with Auto-Correlation for Long-Term Series Forecasting}, 
      author={Haixu Wu and Jiehui Xu and Jianmin Wang and Mingsheng Long},
      year={2022},
      eprint={2106.13008},
      archivePrefix={arXiv},
      primaryClass={cs.LG},
      url={https://arxiv.org/abs/2106.13008}, 
}

@inproceedings{zhou2022fedformer,
  title={{FEDformer}: Frequency enhanced decomposed transformer for long-term series forecasting},
  author={Zhou, Tian and Ma, Ziqing and Wen, Qingsong and Wang, Xue and Sun, Liang and Jin, Rong},
  booktitle={Proc. 39th International Conference on Machine Learning (ICML 2022)},
  location = {Baltimore, Maryland},
  pages={},
  year={2022}
}

@inproceedings{
nie2023patchtst,
title={A Time Series is Worth 64 Words:  Long-term Forecasting with Transformers},
author={Yuqi Nie and Nam H Nguyen and Phanwadee Sinthong and Jayant Kalagnanam},
booktitle={The Eleventh International Conference on Learning Representations },
year={2023},
url={https://openreview.net/forum?id=Jbdc0vTOcol}
}

@inproceedings{wu2023timesnet,
  author = {Wu, Haixu and Hu, Tengge and Liu, Yong and Zhou, Hang and Wang, Jianmin and Long, Mingsheng},
  booktitle = {ICLR},
  ee = {https://openreview.net/forum?id=ju_Uqw384Oq},
  publisher = {OpenReview.net},
  title = {TimesNet: Temporal 2D-Variation Modeling for General Time Series Analysis.},
  url = {http://dblp.uni-trier.de/db/conf/iclr/iclr2023.html#WuHLZ0L23},
  year = 2023
}

@article{lim2021temporal,
title = {Temporal Fusion Transformers for interpretable multi-horizon time series forecasting},
journal = {International Journal of Forecasting},
volume = {37},
number = {4},
pages = {1748-1764},
year = {2021},
issn = {0169-2070},
doi = {https://doi.org/10.1016/j.ijforecast.2021.03.012},
url = {https://www.sciencedirect.com/science/article/pii/S0169207021000637},
author = {Bryan Lim and Sercan Ö. Arık and Nicolas Loeff and Tomas Pfister}
}

@article{salinas2020deepar,
title = {DeepAR: Probabilistic forecasting with autoregressive recurrent networks},
journal = {International Journal of Forecasting},
volume = {36},
number = {3},
pages = {1181-1191},
year = {2020},
issn = {0169-2070},
doi = {https://doi.org/10.1016/j.ijforecast.2019.07.001},
url = {https://www.sciencedirect.com/science/article/pii/S0169207019301888},
author = {David Salinas and Valentin Flunkert and Jan Gasthaus and Tim Januschowski}
}

@article{iacus2024conflict,
author = {Fulvio Attinà and Marcello Carammia and Stefano M. Iacus},
title = {Forecasting change in conflict fatalities with dynamic elastic net},
journal = {International Interactions},
volume = {48},
number = {4},
pages = {649--677},
year = {2022},
publisher = {Routledge},
doi = {10.1080/03050629.2022.2090934},
URL = {https://doi.org/10.1080/03050629.2022.2090934},
eprint = {https://doi.org/10.1080/03050629.2022.2090934}
}

@article{carammia22,
	author = {Carammia, Marcello and Iacus, Stefano Maria and Wilkin, Teddy},
	journal = {Scientific Reports},
	number = {1},
	pages = {1457},
	title = {Forecasting asylum-related migration flows with machine learning and data at scale},
	volume = {12},
	year = {2022}}

@misc{Haodong25,
  author       = {Qi, Haodong and Sirbu, Alina and Momeni, Rahman and others},
  title        = {Modeling Climate-Induced Refugee Migration: An Explainable Machine Learning Approach},
  howpublished = {Preprint, Research Square},
  note         = {Version 1},
  year         = {2024},
  month        = {oct},
  doi          = {10.21203/rs.3.rs-4931065/v1},
  url          = {https://doi.org/10.21203/rs.3.rs-4931065/v1}
}

@inproceedings{devlin2019bert,
  title={BERT: Pre-training of Deep Bidirectional Transformers for Language Understanding},
  author={Jacob Devlin and Ming-Wei Chang and Kenton Lee and Kristina Toutanova},
  booktitle={North American Chapter of the Association for Computational Linguistics},
  year={2019},
  url={https://api.semanticscholar.org/CorpusID:52967399}
}

@inproceedings{radford2019language,
  title={Language Models are Unsupervised Multitask Learners},
  author={Alec Radford and Jeff Wu and Rewon Child and David Luan and Dario Amodei and Ilya Sutskever},
  year={2019},
  url={https://api.semanticscholar.org/CorpusID:160025533},
booktitle={OpenAI}
}

@inproceedings{brown2020language,
 author = {Brown, Tom and Mann, Benjamin and Ryder, Nick and Subbiah, Melanie and Kaplan, Jared D and Dhariwal, Prafulla and Neelakantan, Arvind and Shyam, Pranav and Sastry, Girish and Askell, Amanda and Agarwal, Sandhini and Herbert-Voss, Ariel and Krueger, Gretchen and Henighan, Tom and Child, Rewon and Ramesh, Aditya and Ziegler, Daniel and Wu, Jeffrey and Winter, Clemens and Hesse, Chris and Chen, Mark and Sigler, Eric and Litwin, Mateusz and Gray, Scott and Chess, Benjamin and Clark, Jack and Berner, Christopher and McCandlish, Sam and Radford, Alec and Sutskever, Ilya and Amodei, Dario},
 booktitle = {Advances in Neural Information Processing Systems},
 editor = {H. Larochelle and M. Ranzato and R. Hadsell and M.F. Balcan and H. Lin},
 pages = {1877--1901},
 publisher = {Curran Associates, Inc.},
 title = {Language Models are Few-Shot Learners},
 url = {https://proceedings.neurips.cc/paper_files/paper/2020/file/1457c0d6bfcb4967418bfb8ac142f64a-Paper.pdf},
 volume = {33},
 year = {2020}
}

@misc{hendrycks2016gelu,
      title={Gaussian Error Linear Units (GELUs)}, 
      author={Dan Hendrycks and Kevin Gimpel},
      year={2023},
      eprint={1606.08415},
      archivePrefix={arXiv},
      primaryClass={cs.LG},
      url={https://arxiv.org/abs/1606.08415}, 
}

\clearpage
\section*{Additional Figures}

\clearpage
\section*{Additional Tables}

\begin{table}[H]
\centering
{\tiny
\begin{tabular}{llr}
  \toprule
feature\_name & conf\_region & gate \\ 
  \midrule
ged\_sb\_tlag\_6 & COD-RWA-BDI & 0.4990 \\ 
  ged\_sb\_tlag\_12 & COD-RWA-BDI & 0.4847 \\ 
  ged\_ns\_splag\_1 & COD-RWA-BDI & 0.3907 \\ 
  ged\_sb\_tlag\_7 & COD-RWA-BDI & 0.3863 \\ 
  treelag\_2\_ns & COD-RWA-BDI & 0.3764 \\ 
  ged\_ns\_tlag\_1\_splag\_1 & COD-RWA-BDI & 0.3678 \\ 
  wdi\_nv\_agr\_totl\_kd & COD-RWA-BDI & 0.3587 \\ 
  ged\_ns\_decay\_12\_time\_since & COD-RWA-BDI & 0.3501 \\ 
  decay\_ged\_ns\_25 & COD-RWA-BDI & 0.3480 \\ 
  decay\_ged\_os\_100 & COD-RWA-BDI & 0.3408 \\ 
  ged\_os\_tlag\_1 & COD-RWA-BDI & 0.3358 \\ 
  treelag\_2\_sb & COD-RWA-BDI & 0.3358 \\ 
  decay\_ged\_os\_25 & COD-RWA-BDI & 0.3357 \\ 
  tlag1\_dr\_moder\_gs & COD-RWA-BDI & 0.3347 \\ 
  spei1\_gsm\_cv\_anom & COD-RWA-BDI & 0.3180 \\ 
  treelag\_1\_sb & COD-RWA-BDI & 0.3115 \\ 
  decay\_ged\_os\_500 & COD-RWA-BDI & 0.3083 \\ 
  tlag1\_spei1\_gsm & COD-RWA-BDI & 0.3077 \\ 
  ged\_os\_tlag\_1\_splag\_1 & COD-RWA-BDI & 0.3068 \\ 
  sptime\_dist\_k10\_ged\_os & COD-RWA-BDI & 0.3020 \\ 
  ged\_os\_decay\_12\_time\_since & COD-RWA-BDI & 0.3013 \\ 
  treelag\_1\_ns & COD-RWA-BDI & 0.3012 \\ 
  tlag1\_dr\_mod\_gs & COD-RWA-BDI & 0.2993 \\ 
  decay\_ged\_os\_5 & COD-RWA-BDI & 0.2985 \\ 
  ged\_os\_splag\_1 & COD-RWA-BDI & 0.2956 \\ 
  growseasdummy & COD-RWA-BDI & 0.2924 \\ 
  count\_moder\_drought\_prev10 & COD-RWA-BDI & 0.2909 \\ 
  spei\_48\_detrend & COD-RWA-BDI & 0.2900 \\ 
  ged\_ns\_tlag\_1 & COD-RWA-BDI & 0.2882 \\ 
  sptime\_dist\_k1\_ged\_sb & COD-RWA-BDI & 0.2800 \\ 
  sptime\_dist\_k10\_ged\_sb & COD-RWA-BDI & 0.2797 \\ 
  ged\_sb\_tlag\_1\_splag\_1 & COD-RWA-BDI & 0.2735 \\ 
  sptime\_dist\_k1\_ged\_os & COD-RWA-BDI & 0.2721 \\ 
  tlag1\_dr\_sev\_gs & COD-RWA-BDI & 0.2699 \\ 
  spei1\_gs\_prev10 & COD-RWA-BDI & 0.2482 \\ 
  spei1\_gsm\_detrend & COD-RWA-BDI & 0.2304 \\ 
  sptime\_dist\_k10\_ged\_ns & COD-RWA-BDI & 0.2165 \\ 
  spei1\_gs\_prev10\_anom & COD-RWA-BDI & 0.2141 \\ 
  spei1gsy\_lowermedian\_count & COD-RWA-BDI & 0.2043 \\ 
  pgd\_nlights\_calib\_mean & COD-RWA-BDI & 0.1255 \\ 
  splag\_1\_1\_sb\_1 & COD-RWA-BDI & 0.1213 \\ 
  splag\_1\_decay\_ged\_sb\_1 & COD-RWA-BDI & 0.1140 \\ 
  treelag\_1\_os & COD-RWA-BDI & 0.0020 \\ 
  decay\_ged\_sb\_1 & COD-RWA-BDI & 0.0019 \\ 
  dist\_petroleum & COD-RWA-BDI & 0.0016 \\ 
  pasture\_ih & COD-RWA-BDI & 0.0015 \\ 
  sptime\_dist\_k001\_ged\_sb & COD-RWA-BDI & 0.0014 \\ 
  ln\_pop\_gpw\_sum & COD-RWA-BDI & 0.0014 \\ 
  tlag\_12\_rainf\_maincrops & COD-RWA-BDI & 0.0013 \\ 
  agri\_ih & COD-RWA-BDI & 0.0013 \\ 
  forest\_ih & COD-RWA-BDI & 0.0012 \\ 
  ln\_capdist & COD-RWA-BDI & 0.0011
  \end{tabular}
  }
  \caption{Average gates values for the COD-RWA-BDI conflict system}
  \label{tab:gate_COD_RWA_BDI}
  \end{table}

\begin{table}[H]
\centering
{\tiny 
\begin{tabular}{llr}
  \toprule
feature\_name & conf\_region & gate \\ 
  \midrule
  treelag\_2\_ns & ISR-LBN-SYR & 0.3123 \\ 
  treelag\_1\_sb & ISR-LBN-SYR & 0.3062 \\ 
  ged\_sb\_tlag\_9 & ISR-LBN-SYR & 0.2736 \\ 
  ged\_os\_splag\_1 & ISR-LBN-SYR & 0.2388 \\ 
  ged\_gte\_1 & ISR-LBN-SYR & 0.2366 \\ 
  tlag1\_dr\_mod\_gs & ISR-LBN-SYR & 0.2321 \\ 
  ged\_sb\_tlag\_12 & ISR-LBN-SYR & 0.2263 \\ 
  ged\_os\_tlag\_1 & ISR-LBN-SYR & 0.2186 \\ 
  ged\_sb\_tlag\_11 & ISR-LBN-SYR & 0.2175 \\ 
  ged\_sb\_tlag\_2 & ISR-LBN-SYR & 0.2170 \\ 
  ged\_sb\_tlag\_8 & ISR-LBN-SYR & 0.2165 \\ 
  treelag\_1\_ns & ISR-LBN-SYR & 0.2157 \\ 
  ged\_sb\_tlag\_3 & ISR-LBN-SYR & 0.2130 \\ 
  ged\_sb\_tlag\_4 & ISR-LBN-SYR & 0.2085 \\ 
  sptime\_dist\_k1\_ged\_sb & ISR-LBN-SYR & 0.1953 \\ 
  tlag1\_spei1\_gsm & ISR-LBN-SYR & 0.1920 \\ 
  ged\_os\_tlag\_1\_splag\_1 & ISR-LBN-SYR & 0.1908 \\ 
  ged\_sb\_tlag\_1 & ISR-LBN-SYR & 0.1845 \\ 
  tlag1\_dr\_moder\_gs & ISR-LBN-SYR & 0.1815 \\ 
  growseasdummy & ISR-LBN-SYR & 0.1803 \\ 
  spei1\_gsm\_cv\_anom & ISR-LBN-SYR & 0.1599 \\ 
  sptime\_dist\_k10\_ged\_ns & ISR-LBN-SYR & 0.1585 \\ 
  ged\_sb\_decay\_12\_time\_since & ISR-LBN-SYR & 0.1507 \\ 
  sptime\_dist\_k10\_ged\_sb & ISR-LBN-SYR & 0.1454 \\ 
  spei1\_gsm\_detrend & ISR-LBN-SYR & 0.1282 \\ 
  treelag\_2\_sb & ISR-LBN-SYR & 0.1233 \\ 
  pgd\_imr\_mean & ISR-LBN-SYR & 0.1181 \\ 
  mov\_avg\_12\_ged\_best\_sb & ISR-LBN-SYR & 0.1136 \\ 
  ged\_sb\_tlag\_10 & ISR-LBN-SYR & 0.1113 \\ 
  spei1\_gs\_prev10\_anom & ISR-LBN-SYR & 0.1057 \\ 
  tlag1\_dr\_sev\_gs & ISR-LBN-SYR & 0.1023 \\ 
  mov\_sum\_12\_ged\_best\_sb & ISR-LBN-SYR & 0.0966 \\ 
  ged\_sb\_splag\_1 & ISR-LBN-SYR & 0.0952 \\ 
  imr\_mean & ISR-LBN-SYR & 0.0876 \\ 
  sptime\_dist\_k1\_ged\_ns & ISR-LBN-SYR & 0.0857 \\ 
  ged\_ns\_decay\_12\_time\_since & ISR-LBN-SYR & 0.0815 \\ 
  decay\_ged\_sb\_1 & ISR-LBN-SYR & 0.0779 \\ 
  ged\_sb\_tlag\_1\_splag\_1 & ISR-LBN-SYR & 0.0637 \\ 
  sptime\_dist\_k1\_ged\_os & ISR-LBN-SYR & 0.0608 \\ 
  sptime\_dist\_k10\_ged\_os & ISR-LBN-SYR & 0.0525 \\ 
  sptime\_dist\_k001\_ged\_sb & ISR-LBN-SYR & 0.0421 \\ 
  treelag\_2\_os & ISR-LBN-SYR & 0.0412 \\ 
  spei1gsy\_lowermedian\_count & ISR-LBN-SYR & 0.0411 \\ 
  splag\_1\_1\_sb\_1 & ISR-LBN-SYR & 0.0406 \\ 
  sptime\_dist\_k001\_ged\_os & ISR-LBN-SYR & 0.0371 \\ 
  tlag\_12\_harvarea\_maincrops & ISR-LBN-SYR & 0.0324 \\ 
  ln\_capdist & ISR-LBN-SYR & 0.0305 \\ 
  spei\_48\_detrend & ISR-LBN-SYR & 0.0271 \\ 
  pasture\_ih & ISR-LBN-SYR & 0.0262 \\ 
  pgd\_nlights\_calib\_mean & ISR-LBN-SYR & 0.0256 \\ 
  count\_moder\_drought\_prev10 & ISR-LBN-SYR & 0.0236 \\ 
  splag\_1\_decay\_ged\_sb\_1 & ISR-LBN-SYR & 0.0235 \\ 
  decay\_ged\_sb\_500 & ISR-LBN-SYR & 0.0215 \\ 
  ged\_os\_decay\_12\_time\_since & ISR-LBN-SYR & 0.0207 \\ 
  pgd\_urban\_ih & ISR-LBN-SYR & 0.0181 \\ 
  ln\_bdist3 & ISR-LBN-SYR & 0.0174 \\ 
  tlag\_12\_crop\_sum & ISR-LBN-SYR & 0.0164 \\ 
  ged\_ns\_splag\_1 & ISR-LBN-SYR & 0.0155 \\ 
  barren\_ih & ISR-LBN-SYR & 0.0147 \\ 
  sptime\_dist\_k001\_ged\_ns & ISR-LBN-SYR & 0.0140 \\ 
  wdi\_nv\_agr\_totl\_kd & ISR-LBN-SYR & 0.0138 \\ 
  agri\_ih & ISR-LBN-SYR & 0.0138 \\ 
  ln\_gcp\_mer & ISR-LBN-SYR & 0.0136 \\ 
  shrub\_ih & ISR-LBN-SYR & 0.0128 \\ 
  spei1\_gs\_prev10 & ISR-LBN-SYR & 0.0124 \\ 
  ln\_pop\_gpw\_sum & ISR-LBN-SYR & 0.0118 \\ 
  urban\_ih & ISR-LBN-SYR & 0.0064 \\ 
  forest\_ih & ISR-LBN-SYR & 0.0036 \\ 
  decay\_ged\_ns\_25 & ISR-LBN-SYR & 0.0025 \\ 
  decay\_ged\_os\_5 & ISR-LBN-SYR & 0.0020 \\ 
  decay\_ged\_os\_1 & ISR-LBN-SYR & 0.0020 \\ 
  dist\_diamsec & ISR-LBN-SYR & 0.0015 \\ 
  decay\_ged\_sb\_100 & ISR-LBN-SYR & 0.0010 \\ 
  ln\_ttime\_mean & ISR-LBN-SYR & 0.0010 \\ 
\end{tabular}
}
  \caption{Average gates values for the ISR-LBN-SYR conflict system}
  \label{tab:gate_ISR-LBN-SYR}
\end{table}

\begin{table}[H]
\centering
{\tiny
\begin{tabular}{llr}
  \toprule
feature\_name & conf\_region & gate \\ 
  \midrule
  treelag\_2\_ns & NGA-BFA-MLI & 0.2897 \\ 
  spei1\_gsm\_detrend & NGA-BFA-MLI & 0.2479 \\ 
  ged\_ns\_splag\_1 & NGA-BFA-MLI & 0.2218 \\ 
  spei1\_gs\_prev10\_anom & NGA-BFA-MLI & 0.1977 \\ 
  tlag1\_spei1\_gsm & NGA-BFA-MLI & 0.1976 \\ 
  decay\_ged\_ns\_1 & NGA-BFA-MLI & 0.1945 \\ 
  treelag\_1\_sb & NGA-BFA-MLI & 0.1916 \\ 
  treelag\_2\_sb & NGA-BFA-MLI & 0.1803 \\ 
  growseasdummy & NGA-BFA-MLI & 0.1746 \\ 
  ged\_ns\_tlag\_1\_splag\_1 & NGA-BFA-MLI & 0.1688 \\ 
  spei1\_gsm\_cv\_anom & NGA-BFA-MLI & 0.1378 \\ 
  sptime\_dist\_k1\_ged\_sb & NGA-BFA-MLI & 0.1305 \\ 
  sptime\_dist\_k1\_ged\_os & NGA-BFA-MLI & 0.1295 \\ 
  sptime\_dist\_k10\_ged\_ns & NGA-BFA-MLI & 0.1134 \\ 
  decay\_ged\_ns\_5 & NGA-BFA-MLI & 0.0908 \\ 
  sptime\_dist\_k10\_ged\_sb & NGA-BFA-MLI & 0.0905 \\ 
  treelag\_1\_ns & NGA-BFA-MLI & 0.0725 \\ 
  sptime\_dist\_k001\_ged\_os & NGA-BFA-MLI & 0.0722 \\ 
  tlag1\_dr\_mod\_gs & NGA-BFA-MLI & 0.0687 \\ 
  tlag1\_dr\_moder\_gs & NGA-BFA-MLI & 0.0685 \\ 
  sptime\_dist\_k10\_ged\_os & NGA-BFA-MLI & 0.0603 \\ 
  spei1\_gs\_prev10 & NGA-BFA-MLI & 0.0312 \\ 
  count\_moder\_drought\_prev10 & NGA-BFA-MLI & 0.0304 \\ 
  spei1gsy\_lowermedian\_count & NGA-BFA-MLI & 0.0245 \\ 
  pasture\_ih & NGA-BFA-MLI & 0.0151 \\ 
  ged\_sb\_tlag\_1\_splag\_1 & NGA-BFA-MLI & 0.0104 \\ 
  sptime\_dist\_k001\_ged\_sb & NGA-BFA-MLI & 0.0095 \\ 
  sptime\_dist\_k1\_ged\_ns & NGA-BFA-MLI & 0.0074 \\ 
  ged\_os\_splag\_1 & NGA-BFA-MLI & 0.0063 \\ 
  ln\_pgd\_ttime\_mean & NGA-BFA-MLI & 0.0052 \\ 
  decay\_ged\_sb\_1 & NGA-BFA-MLI & 0.0050 \\ 
  splag\_1\_1\_sb\_1 & NGA-BFA-MLI & 0.0050 \\ 
  tlag\_12\_crop\_sum & NGA-BFA-MLI & 0.0044 \\ 
  tlag\_12\_harvarea\_maincrops & NGA-BFA-MLI & 0.0034 \\ 
  ln\_gcp\_mer & NGA-BFA-MLI & 0.0033 \\ 
  agri\_ih & NGA-BFA-MLI & 0.0025 \\ 
  splag\_1\_decay\_ged\_sb\_1 & NGA-BFA-MLI & 0.0025 \\ 
  tlag1\_dr\_sev\_gs & NGA-BFA-MLI & 0.0025 \\ 
  spei\_48\_detrend & NGA-BFA-MLI & 0.0023 \\ 
  pgd\_urban\_ih & NGA-BFA-MLI & 0.0022 \\ 
  savanna\_ih & NGA-BFA-MLI & 0.0019 \\ 
  sptime\_dist\_k001\_ged\_ns & NGA-BFA-MLI & 0.0018 \\ 
  ln\_pop\_gpw\_sum & NGA-BFA-MLI & 0.0017 \\ 
  decay\_ged\_sb\_5 & NGA-BFA-MLI & 0.0015 \\ 
  ln\_capdist & NGA-BFA-MLI & 0.0015 \\ 
  imr\_mean & NGA-BFA-MLI & 0.0014 \\ 
  pgd\_imr\_mean & NGA-BFA-MLI & 0.0014 \\ 
  pgd\_nlights\_calib\_mean & NGA-BFA-MLI & 0.0013 \\ 
  dist\_diamsec & NGA-BFA-MLI & 0.0013 \\ 
  decay\_ged\_os\_1 & NGA-BFA-MLI & 0.0012 \\ 
  urban\_ih & NGA-BFA-MLI & 0.0012 \\ 
  wdi\_nv\_agr\_totl\_kd & NGA-BFA-MLI & 0.0012 \\ 
  decay\_ged\_ns\_500 & NGA-BFA-MLI & 0.0011 \\ 
  dist\_petroleum & NGA-BFA-MLI & 0.0011 \\ 
  shrub\_ih & NGA-BFA-MLI & 0.0011 \\ 
  tlag\_12\_irr\_maincrops & NGA-BFA-MLI & 0.0008 \\ 
  ln\_bdist3 & NGA-BFA-MLI & 0.0008 \\ 
  ged\_os\_tlag\_1\_splag\_1 & NGA-BFA-MLI & 0.0008 \\ 
  ln\_ttime\_mean & NGA-BFA-MLI & 0.0007 \\ 
  decay\_ged\_ns\_25 & NGA-BFA-MLI & 0.0007 \\ 
  ged\_sb\_decay\_12\_time\_since & NGA-BFA-MLI & 0.0005 \\ 
  decay\_ged\_sb\_25 & NGA-BFA-MLI & 0.0004 \\ 
  forest\_ih & NGA-BFA-MLI & 0.0004 \\ 
  decay\_ged\_os\_5 & NGA-BFA-MLI & 0.0003 \\ 
\end{tabular}
}
  \caption{Average gates values for the NGA-BFA-MLI conflict system}
  \label{tab:gate_NGA-BFA-MLI}
\end{table}
\clearpage

\begin{table}[H]
\centering
{\tiny
\begin{tabular}{llr}
  \toprule
feature\_name & conf\_region & gate \\ 
  \midrule  
  ged\_os\_decay\_12\_time\_since & SDN-ETH-SOM & 0.4119 \\ 
  tlag1\_dr\_moder\_gs & SDN-ETH-SOM & 0.3046 \\ 
  tlag1\_spei1\_gsm & SDN-ETH-SOM & 0.2593 \\ 
  sptime\_dist\_k1\_ged\_sb & SDN-ETH-SOM & 0.2542 \\ 
  tlag1\_dr\_mod\_gs & SDN-ETH-SOM & 0.2516 \\ 
  ged\_ns\_splag\_1 & SDN-ETH-SOM & 0.2478 \\ 
  treelag\_1\_sb & SDN-ETH-SOM & 0.2313 \\ 
  growseasdummy & SDN-ETH-SOM & 0.1973 \\ 
  spei1\_gsm\_detrend & SDN-ETH-SOM & 0.1801 \\ 
  sptime\_dist\_k1\_ged\_os & SDN-ETH-SOM & 0.1754 \\ 
  decay\_ged\_sb\_5 & SDN-ETH-SOM & 0.1721 \\ 
  spei1\_gs\_prev10\_anom & SDN-ETH-SOM & 0.1676 \\ 
  spei1\_gsm\_cv\_anom & SDN-ETH-SOM & 0.1606 \\ 
  sptime\_dist\_k10\_ged\_sb & SDN-ETH-SOM & 0.1542 \\ 
  spei1\_gs\_prev10 & SDN-ETH-SOM & 0.1540 \\ 
  count\_moder\_drought\_prev10 & SDN-ETH-SOM & 0.1523 \\ 
  decay\_ged\_ns\_1 & SDN-ETH-SOM & 0.1438 \\ 
  sptime\_dist\_k10\_ged\_ns & SDN-ETH-SOM & 0.1392 \\ 
  sptime\_dist\_k10\_ged\_os & SDN-ETH-SOM & 0.1369 \\ 
  ged\_os\_splag\_1 & SDN-ETH-SOM & 0.1147 \\ 
  spei\_48\_detrend & SDN-ETH-SOM & 0.1146 \\ 
  treelag\_1\_ns & SDN-ETH-SOM & 0.1098 \\ 
  sptime\_dist\_k1\_ged\_ns & SDN-ETH-SOM & 0.0917 \\ 
  decay\_ged\_sb\_1 & SDN-ETH-SOM & 0.0898 \\ 
  tlag\_12\_irr\_maincrops & SDN-ETH-SOM & 0.0786 \\ 
  tlag\_12\_rainf\_maincrops & SDN-ETH-SOM & 0.0496 \\ 
  dist\_diamsec & SDN-ETH-SOM & 0.0452 \\ 
  splag\_1\_decay\_ged\_sb\_1 & SDN-ETH-SOM & 0.0432 \\ 
  wdi\_nv\_agr\_totl\_kd & SDN-ETH-SOM & 0.0393 \\ 
  spei1gsy\_lowermedian\_count & SDN-ETH-SOM & 0.0379 \\ 
  dist\_petroleum & SDN-ETH-SOM & 0.0324 \\ 
  sptime\_dist\_k001\_ged\_ns & SDN-ETH-SOM & 0.0315 \\ 
  ln\_capdist & SDN-ETH-SOM & 0.0281 \\ 
  splag\_1\_1\_sb\_1 & SDN-ETH-SOM & 0.0268 \\ 
  sptime\_dist\_k001\_ged\_os & SDN-ETH-SOM & 0.0256 \\ 
  pgd\_nlights\_calib\_mean & SDN-ETH-SOM & 0.0243 \\ 
  pasture\_ih & SDN-ETH-SOM & 0.0206 \\ 
  ln\_gcp\_mer & SDN-ETH-SOM & 0.0107 \\ 
  shrub\_ih & SDN-ETH-SOM & 0.0103 \\ 
  agri\_ih & SDN-ETH-SOM & 0.0103 \\ 
  ln\_bdist3 & SDN-ETH-SOM & 0.0102 \\ 
  ln\_pop\_gpw\_sum & SDN-ETH-SOM & 0.0095 \\ 
  decay\_ged\_sb\_25 & SDN-ETH-SOM & 0.0075 \\ 
  tlag\_12\_crop\_sum & SDN-ETH-SOM & 0.0074 \\ 
  tlag\_12\_harvarea\_maincrops & SDN-ETH-SOM & 0.0066 \\ 
  greq\_1\_excluded & SDN-ETH-SOM & 0.0042 \\ 
  ln\_pgd\_ttime\_mean & SDN-ETH-SOM & 0.0023 \\ 
  sptime\_dist\_k001\_ged\_sb & SDN-ETH-SOM & 0.0023 \\ 
  pgd\_urban\_ih & SDN-ETH-SOM & 0.0023 \\ 
  urban\_ih & SDN-ETH-SOM & 0.0019 \\ 
  barren\_ih & SDN-ETH-SOM & 0.0015 \\ 
  ln\_ttime\_mean & SDN-ETH-SOM & 0.0011 \\ 
  pgd\_imr\_mean & SDN-ETH-SOM & 0.0010 \\ 
  decay\_ged\_os\_5 & SDN-ETH-SOM & 0.0010 \\ 
  decay\_ged\_ns\_25 & SDN-ETH-SOM & 0.0009 \\ 
  mov\_sum\_36\_ged\_best\_sb & SDN-ETH-SOM & 0.0009 \\ 
  imr\_mean & SDN-ETH-SOM & 0.0009 \\ 
  savanna\_ih & SDN-ETH-SOM & 0.0008 \\ 
  forest\_ih & SDN-ETH-SOM & 0.0006 \\ 
  cropprop & SDN-ETH-SOM & 0.0005 \\ 
  decay\_ged\_os\_1 & SDN-ETH-SOM & 0.0005 \\ 
  ged\_os\_tlag\_1\_splag\_1 & SDN-ETH-SOM & 0.0003 \\ 
  ged\_sb\_decay\_12\_time\_since & SDN-ETH-SOM & 0.0003 \\ 
  decay\_ged\_sb\_100 & SDN-ETH-SOM & 0.0003 \\ 
  mov\_sum\_6\_ged\_best\_sb & SDN-ETH-SOM & 0.0002 \\ 
\end{tabular}
}
  \caption{Average gates values for the SDN-ETH-SOM conflict system}
  \label{tab:gate_SDN_ETH_SOM}
\end{table}

\clearpage

\section*{Supplementary Material}
\renewcommand{\thefigure}{S\arabic{figure}}
\renewcommand{\thetable}{S\arabic{table}}
\renewcommand{\theequation}{S\arabic{equation}}
\setcounter{figure}{0}
\setcounter{table}{0}
\setcounter{equation}{0}

\subsection*{Additional figures}

\begin{figure}[]
    \caption{Log difference between predicted and true values, country-level}
    \label{fig:cm_err_map}
    \centering
    \includegraphics[width=1\linewidth]{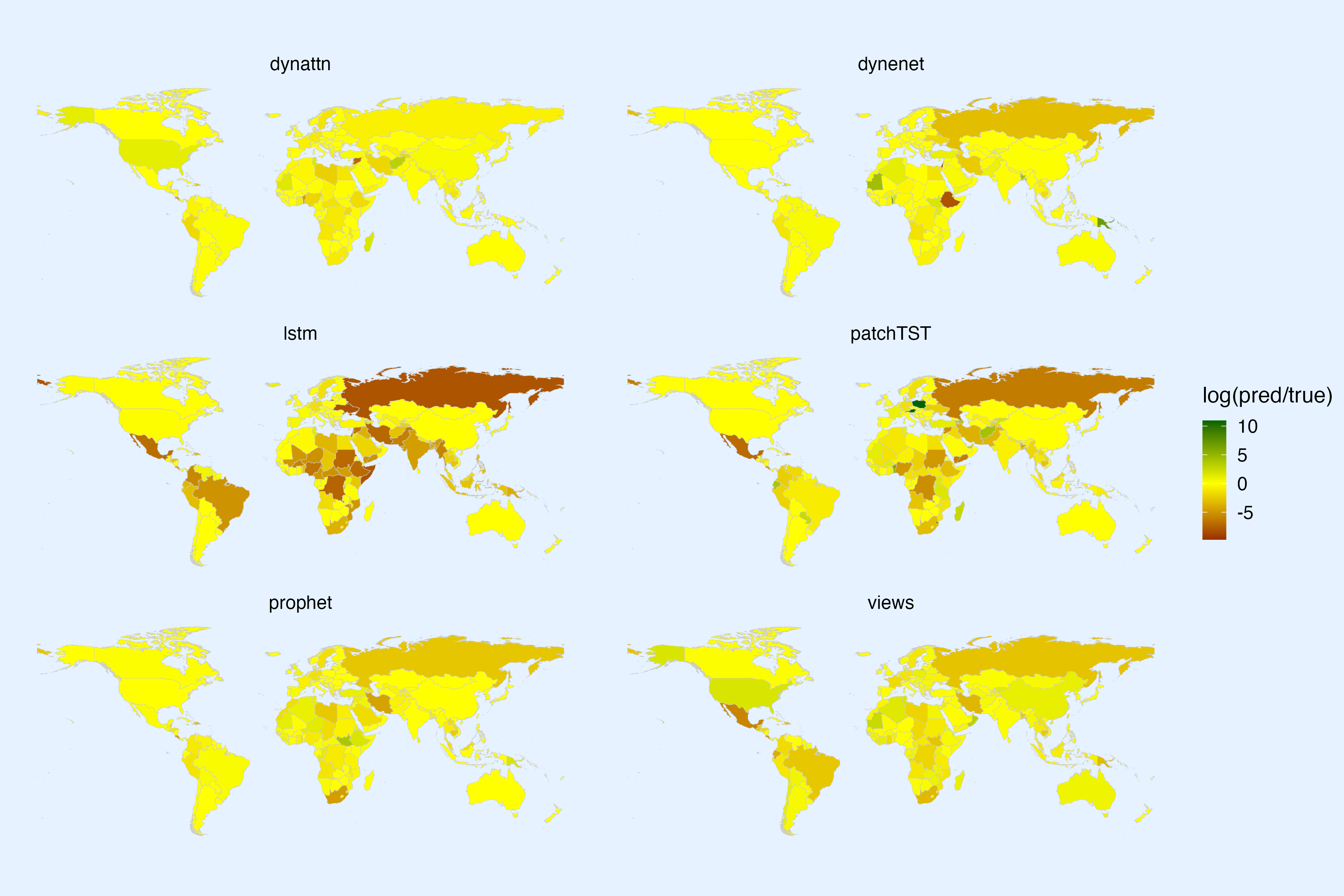}
\end{figure}

\begin{figure}[]
    \caption{Log difference between predicted and true values, grid-level}
    \label{fig:pgm_err_map}
    \centering
    \includegraphics[width=1\linewidth]{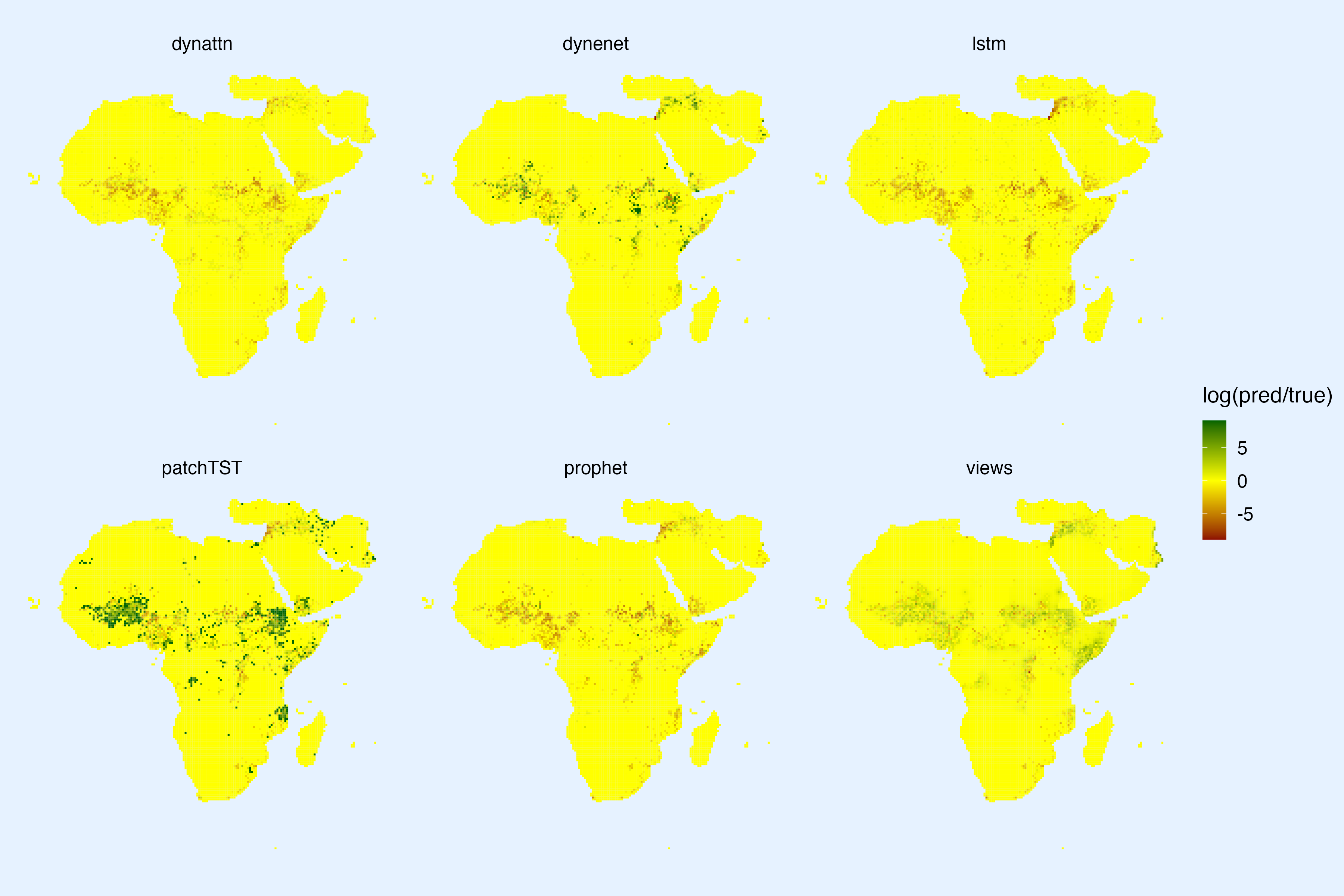}
\end{figure}

\end{document}